\documentclass[reprint,showkeys,amsmath,amssymb,aps,pre]{revtex4-1}

\usepackage{graphicx}
\usepackage{hyperref}

\newtheorem{theorem}{Theorem}

\DeclareMathOperator{\arcsec}{arcsec}
\AtBeginDocument{\renewcommand{\d}{\textrm{d}}}

\begin{document}

\title{Analytical representation of Gaussian processes in the $\mathcal{A}-\mathcal{T}$ plane}
\thanks{Dedicated to my Mother for her 65th birthday.}

\author{Mariusz Tarnopolski}
\email{mariusz.tarnopolski@uj.edu.pl}
\affiliation{Astronomical Observatory, Jagiellonian University, Orla 171, PL-30-244 Krak\'ow, Poland}

\date{\today}

\begin{abstract}
Closed-form expressions, parametrized by the Hurst exponent $H$ and the length $n$ of a time series, are derived for paths of fractional Brownian motion (fBm) and fractional Gaussian noise (fGn) in the $\mathcal{A}-\mathcal{T}$ plane, composed of the fraction of turning points $\mathcal{T}$ and the Abbe value $\mathcal{A}$. The exact formula for $\mathcal{A}_{\rm fBm}$ is expressed via Riemann $\zeta$ and Hurwitz $\zeta$ functions. A very accurate approximation, yielding a simple exponential form, is obtained. Finite-size effects, introduced by the deviation of fGn's variance from unity, and asymptotic cases are discussed. Expressions for $\mathcal{T}$ for fBm, fGn, and differentiated fGn are also presented. The same methodology, valid for any Gaussian process, is applied to autoregressive moving average processes, for which regions of availability of the $\mathcal{A}-\mathcal{T}$ plane are derived and given in analytic form. Locations in the $\mathcal{A}-\mathcal{T}$ plane of some real-world examples as well as generated data are discussed for illustration.
\end{abstract}

\maketitle

\section{Introduction}
\label{sect1}

The characterization and classification of time series \cite{beran,brockwell,brockwell2,beran2} is an important task in a variety of fields. Several methods have been developed for tasks such as detecting chaos \cite{hegger99}, measuring complexity via entropies \cite{bandt_pompe_02,maggs13,ribeiro17}, estimating the Hurst exponent \cite{simonsen,carbone04,arianos07}, distinguishing chaotic from stochastic processes based on graph theory \cite{lacasa10}, and more. A connection between chaos and long range dependence was also established for low-dimensional chaotic maps \cite{tarnopolski18}.

The $\mathcal{A}-\mathcal{T}$ plane \cite{tarnopolski16}, spanned by the fraction of turning points $\mathcal{T}$ and the Abbe value $\mathcal{A}$, was initially introduced to provide a fast and simple estimate of the Hurst exponent $H$, as tight relations of both $\mathcal{A}$ and $\mathcal{T}$ with $H$ were discovered for fractional Brownian motion (fBm), fractional Gaussian noise (fGn), and differentiated fGn (DfGn). While $\mathcal{A}(H)$ and $\mathcal{T}(H)$ strongly overlap for $H\in(0,1)$ for different processes, in the joint space $(\mathcal{A}, \mathcal{T})$ the fBm and fGn intersect only at the point corresponding to white noise, i.e. $(1, 2/3)$. A few real-world data sets (monthly mean of the sunspot number (SSN); stock market indices; chaotic time series from the Lorenz system and the Chirikov map) were shown to lie firmly on the fBm branch. Moreover, the estimates of $H$ based on the empirical relation $\mathcal{A}(H)$ and computed using a wavelet method \citep{peng94,peng95} were consistent with each other.

The discriminative power of the $\mathcal{A}-\mathcal{T}$ plane was demonstrated in a multiscale scheme \cite{zunino17} by employing coarse-grained sequences, i.e. dividing the time series into nonoverlapping segments of length $\tau$ and calculating the mean in each segment. This produces smoothed sequences, and the evolution of $\mathcal{A}$ and $\mathcal{T}$ with varying temporal scale $\tau$ allowed to separate (i) developed, emerging and frontier stock markets; (ii) healthy and epileptic patients based on their EEG recordings---moreover, for the first time it was possible to distinguish healthy patients with closed and open eyes; and (iii) patients with and without cardiac diseases based on the heart rate variabilty. Finally, it is also possible to differentiate between chaotic and stochastic processes based on the different behavior of paths, parametrized by $\tau$, in the $\mathcal{A}-\mathcal{T}$ plane \cite{zhao18}. The $\mathcal{A}-\mathcal{T}$ plane is therefore a powerful tool with several possible applications.

The aim of this paper is to derive analytical descriptions of fBm and fGn, as well as autoregressive moving average (ARMA) processes, in the $\mathcal{A}-\mathcal{T}$ plane. In Sect.~\ref{sect2.1} the known facts about turning points are recapitulated. In Sect.~\ref{sect2.2} the exact and approximated expressions for $\mathcal{A}$ are derived, for the first time, for fBm and fGn. Their $\mathcal{A}-\mathcal{T}$ plane's representation is depicted in Sect.~\ref{sect2.3}. ARMA processes are discussed in Sect.~\ref{sect2.4}. Various applications, ranging from pure mathematics, to biology and astrophysics, to financial markets, are briefly outlined in Sect.~\ref{sect::applic}. Summary and concluding remarks with future prospects are gathered in Sect.~\ref{sect3}.

\section{Fraction of turning points, $\mathcal{T}$}
\label{sect2.1}

\subsection{Theory}
\label{sect::theory}

\begin{figure}
\centering
\includegraphics[width=0.7\columnwidth]{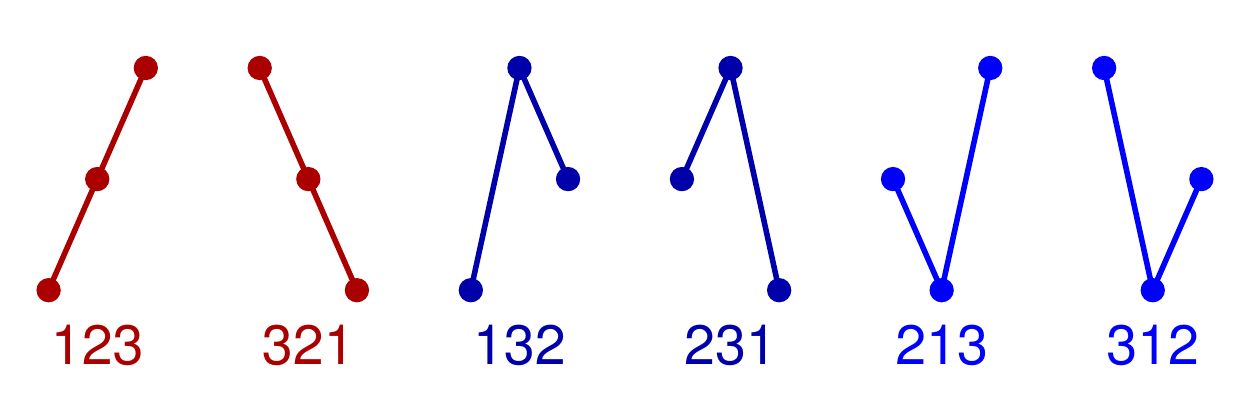}
\caption{Order patterns for three points.}
\label{fig-1}
\end{figure}

Consider three values $x_t, x_{t+d}, x_{t+2d}$ of a time series $\{x_t\}$. For $d=1$ the points are consecutive. Assume there are no ties between the neighboring points, which for continuous processes, or empirical data with decent resolution, should not be an issue (see also \citep{zunino2017,traversaro18}). Three points can be arranged in six ways, identified by an order pattern $\pi_p$ (Fig.~\ref{fig-1}). If the smallest value among the three is given an index 1, and the largest an index 3, then e.g. the relation for one of the four possible turning points, $x_t<x_{t+2d}<x_{t+d}$, is described by a pattern $\pi_p=132$. Denote the probability of encountering a pattern $\pi_p$ by $p_{\pi_p}$. Then the following theorem holds \cite{bandt07}:
\begin{theorem}
For a Gaussian process $X_t$ with stationary increments, $p_{123} = p_{321} = \alpha/2$, and the other patterns yield probability $(1-\alpha)/4$.
\label{theor1}
\end{theorem}
The probability $p_{123}(d)$ for a given delay $d$ is given by
\begin{equation}
p_{123}(d) = \frac{1}{\pi}\arcsin\sqrt{\frac{1+\rho (d)}{2}},
\label{eq36}
\end{equation}
where the correlation coefficient is
\begin{equation}
\rho (d) = \frac{E\left[ \left(X_d-X_0\right)\left(X_{2d}-X_d\right) \right]}{E\left[ \left(X_d-X_0\right)^2 \right]}.
\label{eq37}
\end{equation}
The probability $\mathcal{T}$ of encountering a turning point among three consecutive points (i.e., for $d=1$, the case considered hereinafter), as per Theorem~\ref{theor1}, is then
\begin{equation}
\mathcal{T} = 1 - 2p_{123}(1).
\label{eq38}
\end{equation}
Note that \mbox{$\mathcal{T}\in [0,1]$:} zero is attained by monotonic sequences, while unity is (asymptotically) achieved for strictly alternating time series. For an uncorrelated process all patterns are equally probable, hence $\mathcal{T} = 2/3$. Further details on order patterns can be found in \cite{bandt07}.

\subsection{$\mathcal{T}$ for fBm, fGn, and DfGn}

With the following properties of fBm $B(t)$, fGn $G(t)$, and DfGn $Y(t)$, one can utilize the methodology from Sect.~\ref{sect::theory} to calculate $\mathcal{T}$ for them as a function of $H$:
\begin{subequations}
\begin{align}
E\left[ B^2(t) \right] =& t^{2H},\label{var1}\\
E\left[ B(t)B(s) \right] =& \frac{1}{2}\left( t^{2H} + s^{2H} - |t-s|^{2H} \right),\label{var2}\\
E\left[ G^2(t) \right] =& 1,\label{var3}\\
\begin{split}
E\left[ G(t)G(s) \right] =& \frac{1}{2}\left( |t-s-1|^{2H} + |t-s+1|^{2H}\right) \\
 &- |t-s|^{2H},
\end{split}\label{var4}\\
E\left[ Y^2(t) \right] =& 4-4^H,\label{var5}\\
\begin{split}
E\left[ Y(t)Y(s) \right] =& -3|t-s|^{2H} \\
&+2\left( |t-s-1|^{2H} + |t-s+1|^{2H} \right) \\
&-\frac{1}{2}\left( |t-s-2|^{2H} + |t-s+2|^{2H} \right).
\end{split}\label{var6}
\end{align}
\label{var_formulae}%
\end{subequations}
Eq.~(\ref{var3}) and (\ref{var4}) can be obtained from Eq.~(\ref{var1}) and (\ref{var2}) by substituting $G(t) = B(t+1) - B(t)$; likewise, Eq.~(\ref{var5}) and (\ref{var6}) follow from Eq.~(\ref{var3}) and (\ref{var4}) by using $Y(t) = G(t+1) - G(t)$. \footnote{In the derivation of $\mathcal{T}_{\rm fGn}$ and $\mathcal{T}_{\rm DfGn}$ one can actually employ only the relations for fBm; the calculations would then become quite burdensome, though.}

\begin{figure}
\centering
\includegraphics[width=0.8\columnwidth]{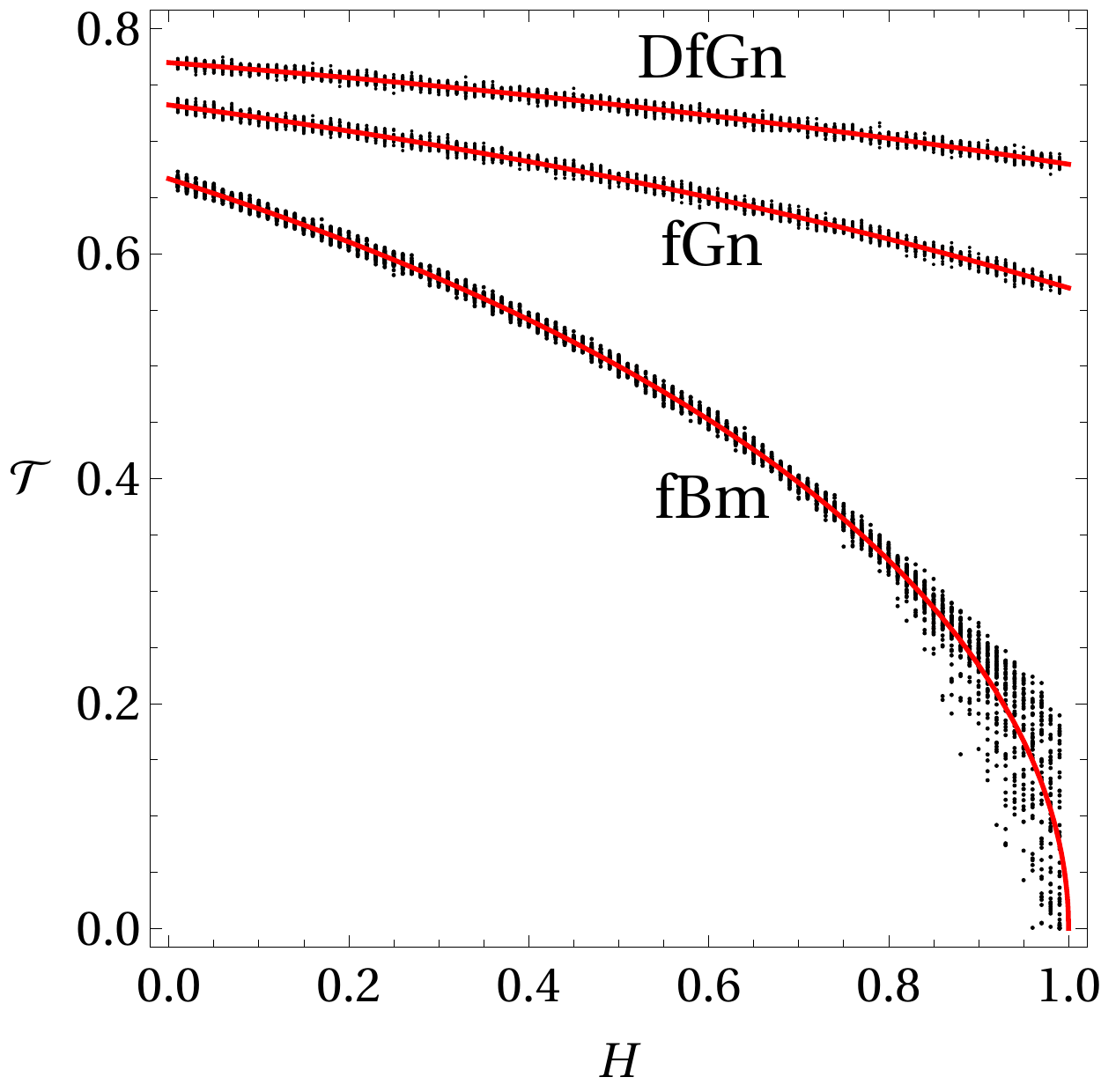}
\caption{Fraction of turning points $\mathcal{T}$ for fBm, fGn, and DfGn (red lines). The simulations were performed for $n=2^{14}$ (black points). The dispersion in case of fBm increases for $H\gtrsim 0.8$ (see \citep{sinn08} for an approximate treatment of the variance of $\mathcal{T}$).}
\label{fig5}
\end{figure}

For fBm, one has
\begin{equation}
\mathcal{T}_{\rm fBm} = 1-\frac{2}{\pi}\arcsin \left( 2^{H-1} \right),
\label{eq1}
\end{equation}
which is roughly equal to $\frac{2}{3}\frac{T}{\mu_T}$ from \cite{tarnopolski16}, where $T$ is the number of turning points in a time series of length $n$ \cite{brockwell,kendall}, and $\mu_T = \frac{2}{3}(n-2)$ is the expected value for white noise \footnote{The first and last points cannot form a turning point, hence the subtraction of 2 in $\mu_T$.}. In general, $E[T]=\left(n-2\right)\mathcal{T}$. The plot of Eq.~(\ref{eq1}) is shown in Fig.~\ref{fig5}, together with data simulated as in \cite{tarnopolski16} (scaled herein from $T/\mu_T$ to $\mathcal{T}$). For an fGn:
\begin{equation}
\mathcal{T}_{\rm fGn} = 1-\frac{2}{\pi}\arcsin\left( \frac{1}{2}\sqrt{\frac{3^{2H}-2^{2H+1}-1}{2^{2H}-4}} \right),
\label{eq32}
\end{equation} 
and for the increments of fGn, i.e. DfGn:
\begin{equation}
\mathcal{T}_{\rm DfGn} = 1-\frac{2}{\pi}\arcsin\left( \frac{1}{2}\sqrt{\frac{(2^{2H}+2)^2-(2\cdot 3^H)^2}{3^{2H}-3\cdot 2^{2H+1}+15}} \right),
\label{eq33}
\end{equation}
both of which are also shown in Fig.~\ref{fig5}.

\section{Abbe value, $\mathcal{A}$}
\label{sect2.2}

The Abbe value of a time series $\{x_i\}_{i=1}^n$ is defined as half the ratio of the mean-square successive difference to the variance \cite{tarnopolski16,neumann,neumann2,kendall1971,mowlavi}:
\begin{equation}
\mathcal{A} = \frac{\frac{1}{n-1}\sum\limits_{i=1}^{n-1}\left( x_{i+1}-x_i \right)^2}{\frac{2}{n}\sum\limits_{i=1}^n\left( x_i-\bar{x} \right)^2}.
\label{eq2}
\end{equation}
It quantifies the smoothness (raggedness) of a time series by comparing the sum of the squared differences between two successive measurements with the variance of the whole time series. It decreases to zero for time series displaying a high degree of smoothness, while the normalization factor ensures that $\mathcal{A}$ tends to unity for a white noise process \cite{williams}. It was proposed as a test for randomness \citep{bingham81,bartels82,mateus18}. It is straightforward to show that for a 2-periodic time series, $\{a,b,a,b,\ldots\}$, $\mathcal{A}=2$, while for a 3-periodic one, $\{a,b,c,a,b,c,\ldots\}$, $\mathcal{A}=3/2$.

Consider $\{x_i\}_{i=1}^n$ to be a realization of length $n$ of an fBm, $B_n^H$, with Hurst exponent $H$. Thence, its increments $\{g_i\}_{i=1}^{n-1}\equiv \{x_{i+1}-x_i\}_{i=1}^{n-1}$ form an fGn, $G_{n-1}^H$, with the same $H$. One can then express $\mathcal{A}$ as
\begin{equation}
\mathcal{A}_{\rm fBm}(H,n) = \frac{1}{2}\frac{{\rm var}\left( G_{n-1}^H \right)}{{\rm var}\left( B_n^H \right)},
\label{eq3}
\end{equation}
where the dependence on $H$ and $n$ is highlighted. Similarly, for an fGn
\begin{equation}
\mathcal{A}_{\rm fGn}(H,n) = \frac{1}{2}\frac{{\rm var}\left( Y_{n-1}^H \right)}{{\rm var}\left( G_n^H \right)},
\label{eq21}
\end{equation}
where $Y_{n-1}^H$ is the corresponding DfGn, i.e. the increments of fGn, $\{y_i\}_{i=1}^{n-1}\equiv \{g_{i+1}-g_i\}_{i=1}^{n-1}$. The goal is to calculate the variances of $B_n^H$, $G_n^H$, and $Y_n^H$.

\subsection{Variance of fGn}
\label{sect2.2.1}

\begin{figure}
\centering
\includegraphics[width=0.8\columnwidth]{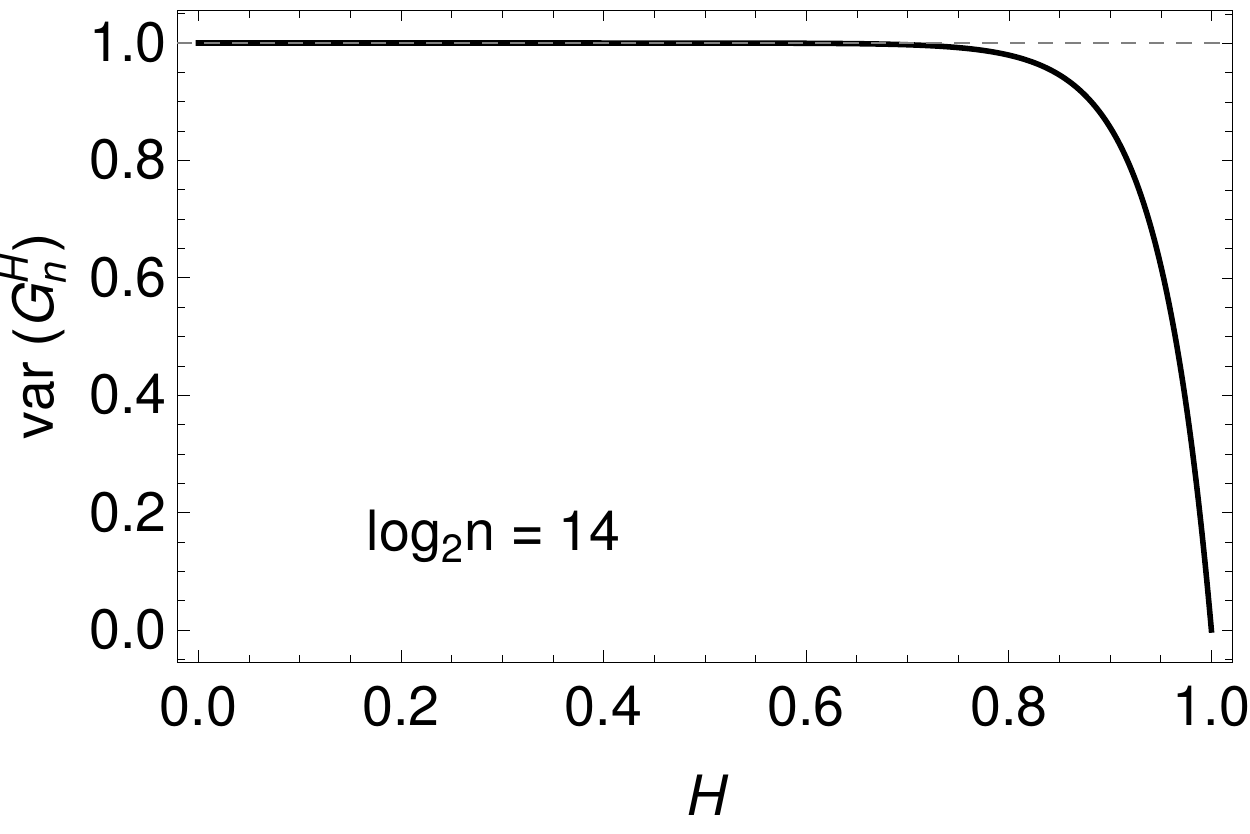}
\caption{Deviation of ${\rm var}\left( G_n^H \right)$ from unity.}
\label{fig0}
\end{figure}

We start with ${\rm var}\left( G_n^H \right)$ since it occurs in both Eq.~(\ref{eq3}) and (\ref{eq21}). The same methodology that was used in \cite{delignieres15} to calculate ${\rm var}\left( B_n^H \right)$ is employed, i.e.
\begin{align}
\begin{split}
{\rm var}\left( G_n^H \right) &= \frac{n}{n-1} E\left[ \left( G_n^H - E\left[ G_n^H \right] \right)^2 \right] \\
&= \frac{1}{n-1} E\left[ \sum\limits_{j=0}^{n-1}\left( G_j - \frac{\sum\limits_{k=0}^{n-1}G_k}{n} \right)^2 \right].
\end{split}
\label{eq25}
\end{align}
Developing the sum for a few values of $n$, and utilizng the variance and covariance from Eq.~(\ref{var3}) and (\ref{var4}), one can observe a pattern emerging, that leads to a formula:
\begin{equation}
{\rm var}\left( G_n^H \right) = \frac{n-n^{2H-1}}{n-1}.
\label{eq26}
\end{equation}
It yields
\begin{equation}
\lim_{n\to\infty}{\rm var}\left( G_n^H \right) \rightarrow 1,
\label{eq27}
\end{equation}
i.e. it asymptotically approaches Eq.~(\ref{var3}). However, for finite $n$, $\lim_{H\to 1}{\rm var}\left( G_n^H \right) = 0$. The plot of Eq.~(\ref{eq26}) for $n=2^{14}$ is shown in Fig.~\ref{fig0}. For values $H\gtrsim 0.8$ the departure from the asymptotic value becomes significant.

\subsection{Variance and Abbe value of fBm}
\label{sect2.2.2}

The variance of the discrete, finite length $B_n^H$ is given in \cite{delignieres15} as
\begin{equation}
{\rm var}\left( B_n^H \right) = \frac{1}{n(n-1)}\sum\limits_{i=1}^{n-1}(n-i)i^{2H}.
\label{eq5}
\end{equation}
For $H\rightarrow 0$, $i^{2H}\rightarrow 1$; then $\sum\limits_{i=1}^{n-1}(n-i) = \frac{n(n-1)}{2}$, so ${\rm var}\left( B_n^0 \right) = \frac{1}{2}$, hence, per Eq.~(\ref{eq3}) and given Eq.~(\ref{eq26}), $\mathcal{A}_{\rm fBm}(H=0,n) = n/(n-1)$, i.e. asymptotically approaches unity.

Using the symbolic computer algebra system \textsc{Mathematica} one can calculate the sum in Eq.~(\ref{eq5}) to be \footnote{See also \url{www.wolframalpha.com}.}
\begin{equation}
\begin{split}
\sum\limits_{i=1}^{n-1}(n-i)i^{2H} =& \zeta(-2H-1,n)-n\zeta(-2H,n)+ \\
& n\zeta(-2H)-\zeta(-2H-1),
\end{split}
\label{eq6}
\end{equation}
where $\zeta(s)$ is the Riemann $\zeta$, and $\zeta(s,n)$ is the Hurwitz $\zeta$ \cite{apostol}. Hence, taking into account Eq.~(\ref{eq26}), one can give a closed-form formula:
\begin{widetext}
\begin{equation}
\mathcal{A}_{\rm fBm}(H,n) = \frac{\frac{n(n-1)}{2}{\rm var}\left( G_{n-1}^H \right)}{\zeta(-2H-1,n)-n\zeta(-2H,n)+n\zeta(-2H)-\zeta(-2H-1)},
\label{eq7}
\end{equation}
\end{widetext}
which, since $\zeta(0,n)=\frac{1}{2}-n$, $\zeta(-1,n) = \frac{1}{2}\left( -\frac{1}{6} +n-n^2 \right)$, $\zeta(0)=-\frac{1}{2}$, and $\zeta(-1)=-\frac{1}{12}$, yields $\mathcal{A}_{\rm fBm}(H=0,n)=n/(n-1)$.

\begin{figure}
\includegraphics[width=0.49\columnwidth]{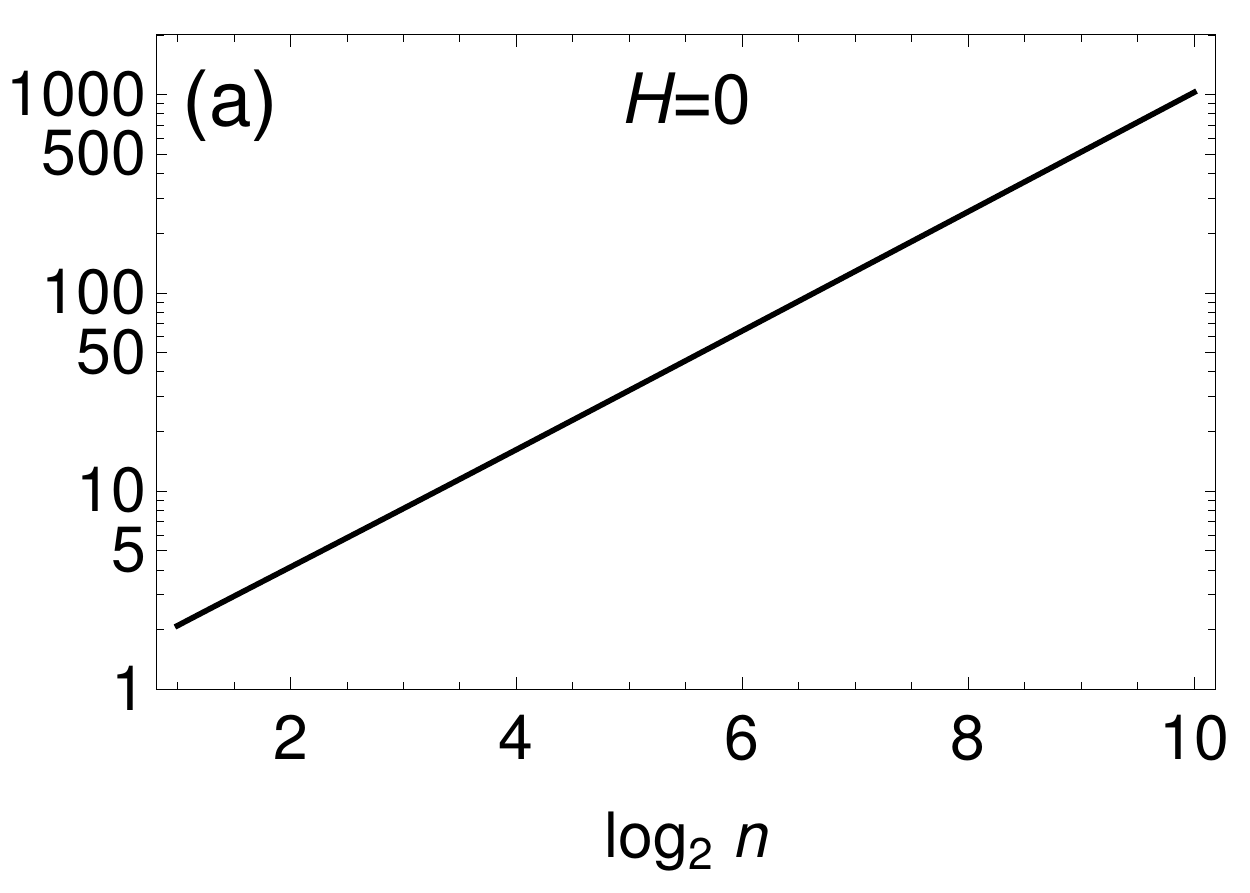}
\includegraphics[width=0.49\columnwidth]{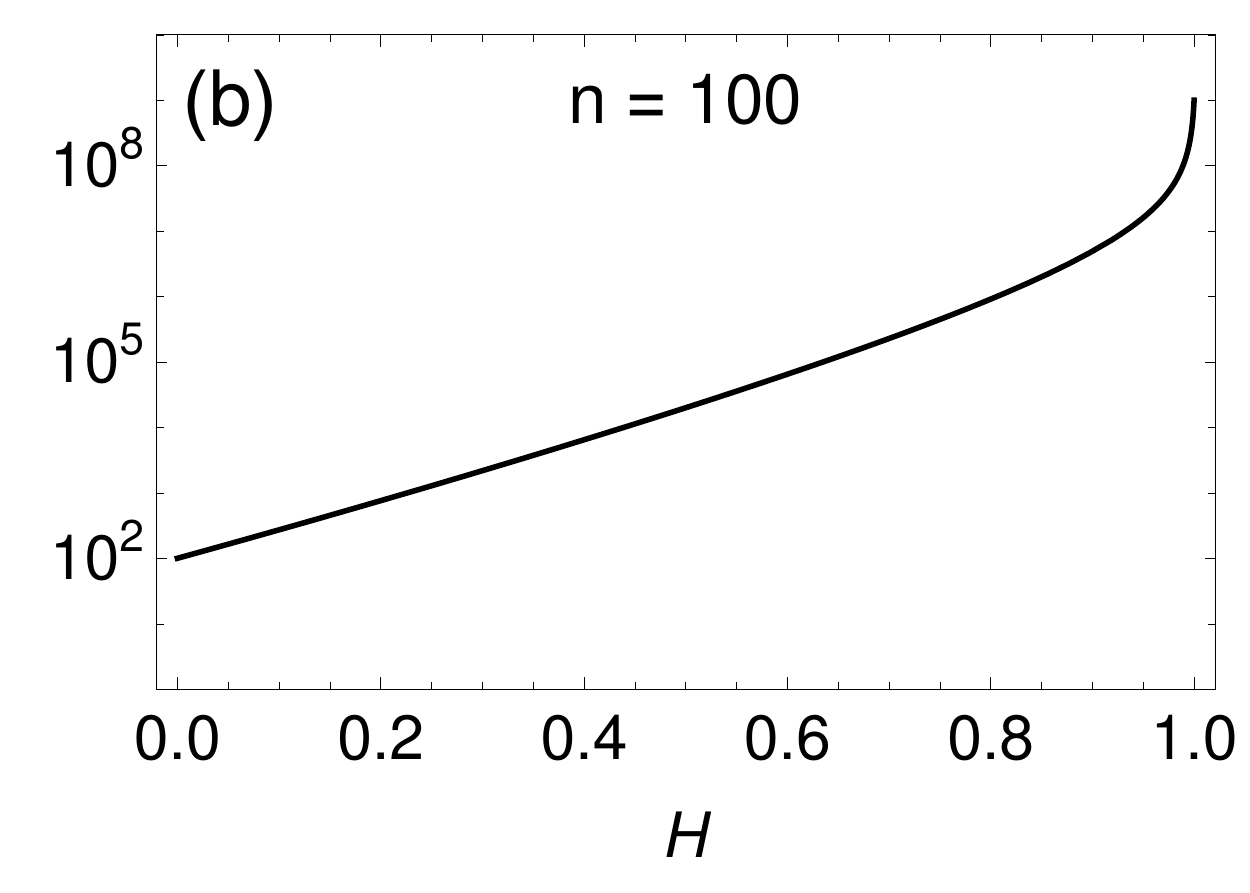}
\caption{The ratio from Eq.~(\ref{eq9}) at (a) $H=0$ for varying $n$, and (b) its dependence on $H$ for a set $n=100$.}
\label{fig1}
\end{figure}

In order to provide a simpler, approximate expression for $\mathcal{A}_{\rm fBm}(H,n)$, first observe that since at $H=0$ and for $n\gg 1$ the ratio of the Hurwitz $\zeta$ terms and the Riemann $\zeta$ terms is big, i.e.
\begin{equation}
\frac{\zeta(-1,n)-n\zeta(0,n)}{|n\zeta(0)-\zeta(-1)|} = \frac{-1+6n^2}{|1-6n|} \gg 1,
\label{eq8}
\end{equation}
and that for $H\in (0,1)$ this ratio is monotonically increasing (Fig.~\ref{fig1}), therefore
\begin{equation}
\frac{\zeta(-2H-1,n)-n\zeta(-2H,n)}{|n\zeta(-2H)-\zeta(-2H-1)|}\gg 1,
\label{eq9}
\end{equation}
hence $n\zeta(-2H)-\zeta(-2H-1)$ is a negligible contribution, so it does not need to be taken into account.

Let us express $\zeta(s,n)$ as a globally convergent Newton series, i.e. utilize the Hasse representation \cite{hasse30}:
\begin{equation}
\zeta(s,n) = \frac{1}{s-1}\sum\limits_{i=0}^\infty\frac{1}{i+1}\sum\limits_{k=0}^i\left(-1\right)^k {i\choose k}\left(n+k\right)^{1-s},
\label{eq10}
\end{equation}
valid for $s\neq 1$, $n>0$. The first term of the outer sum, i.e. for $i=0$ and at $s=-2H$, is
\begin{equation}
\sum\limits_{k=0}^0\left( -1 \right)^k{0\choose k}(n+k)^{1+2H} = n^{1+2H}.
\label{eq11}
\end{equation}
The second term, i.e. for $i=1$, is
\begin{equation}
\frac{1}{2}\sum\limits_{k=0}^1\left( -1 \right)^k{1\choose k}(n+k)^{1+2H} = \frac{1}{2}\left( n^{1+2H} - (n+1)^{1+2H} \right).
\label{eq12}
\end{equation}
And similarly for higher $i$. Therefore, for $i=0$ one obtains an approximation
\begin{equation}
\zeta(-2H,n)\approx \frac{n^{1+2H}}{-2H-1},
\label{eq13}
\end{equation}
and for $i=1$:
\begin{equation}
\zeta(-2H,n)\approx\frac{1}{-2H-1}\left(\frac{3}{2}n^{1+2H}-\frac{1}{2}(n+1)^{1+2H}\right),
\label{eq14}
\end{equation}
but since $n\gg1$, $(n+1)\approx n$, so one also obtains
\begin{equation}
\zeta(-2H,n)\approx \frac{n^{1+2H}}{-2H-1}.
\label{eq15}
\end{equation}
For higher $i$, although given that $i\ll n$, one has $(n+k)\approx n$, hence yielding the same approximation. Indeed, $\sum\limits_{k=0}^i \left(-1\right)^k{i\choose k}$ is the Kronecker $\delta$, $\delta_{i0}$, hence only the $i=0$ term survives. For $s=-2H-1$ one obtains a similar expression:
\begin{equation}
\zeta(-2H-1,n)\approx \frac{n^{2+2H}}{-2H-2}.
\label{eq16}
\end{equation}

Therefore, with these approximations one can write:
\begin{equation}
\begin{split}
\zeta(-2H-1,n)-n\zeta(-2H,n) & \approx \frac{n^{2+2H}}{-2H-2} - \frac{nn^{1+2H}}{-2H-1} \\
& = \frac{n^{2+2H}}{2(H+1)(2H+1)},
\end{split}
\label{eq17}
\end{equation}
so that
\begin{equation}
{\rm var}\left( B_n^H \right) \approx \frac{n^{2+2H}}{2(H+1)(2H+1)n(n-1)},
\label{eq34}
\end{equation}
and since $(n-1)\approx n$, one finally obtains
\begin{equation}
{\rm var}\left( B_n^H \right) \approx \frac{n^{2H}}{2(H+1)(2H+1)}
\label{eq35}
\end{equation}
and
\begin{equation}
\mathcal{A}_{\rm fBm}(H,n) \approx (H+1)(2H+1)n^{-2H} {\rm var}\left( G_{n-1}^H \right),
\label{eq19}
\end{equation}
which also yields $\mathcal{A}_{\rm fBm}(H=0,n)=n/(n-1)$, asymptotically approaching unity. In case one sets ${\rm var}\left( G_{n-1}^H \right) = 1$, the simplest approximation is then obtained as
\begin{equation}
\mathcal{A}_{\rm fBm}(H,n) \approx (H+1)(2H+1)n^{-2H}.
\label{eq30}
\end{equation}

\begin{figure*}
\includegraphics[width=0.66\columnwidth]{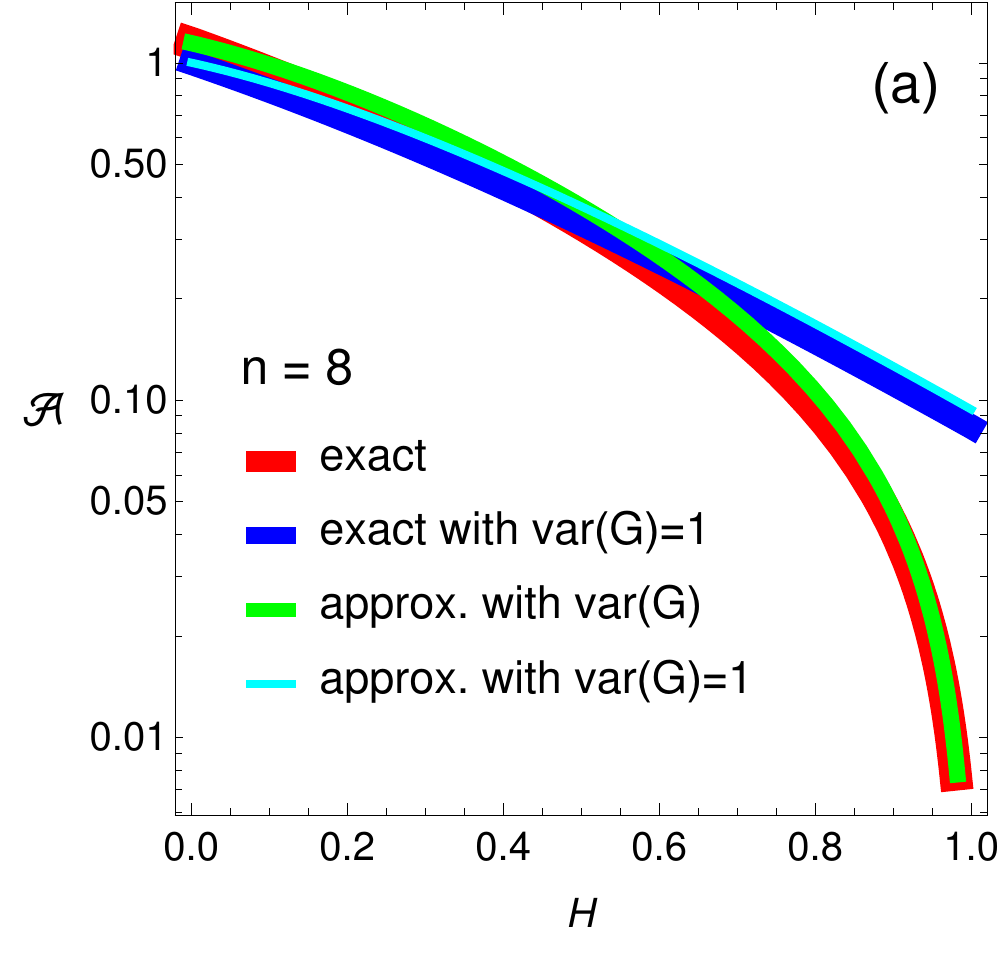}
\includegraphics[width=0.66\columnwidth]{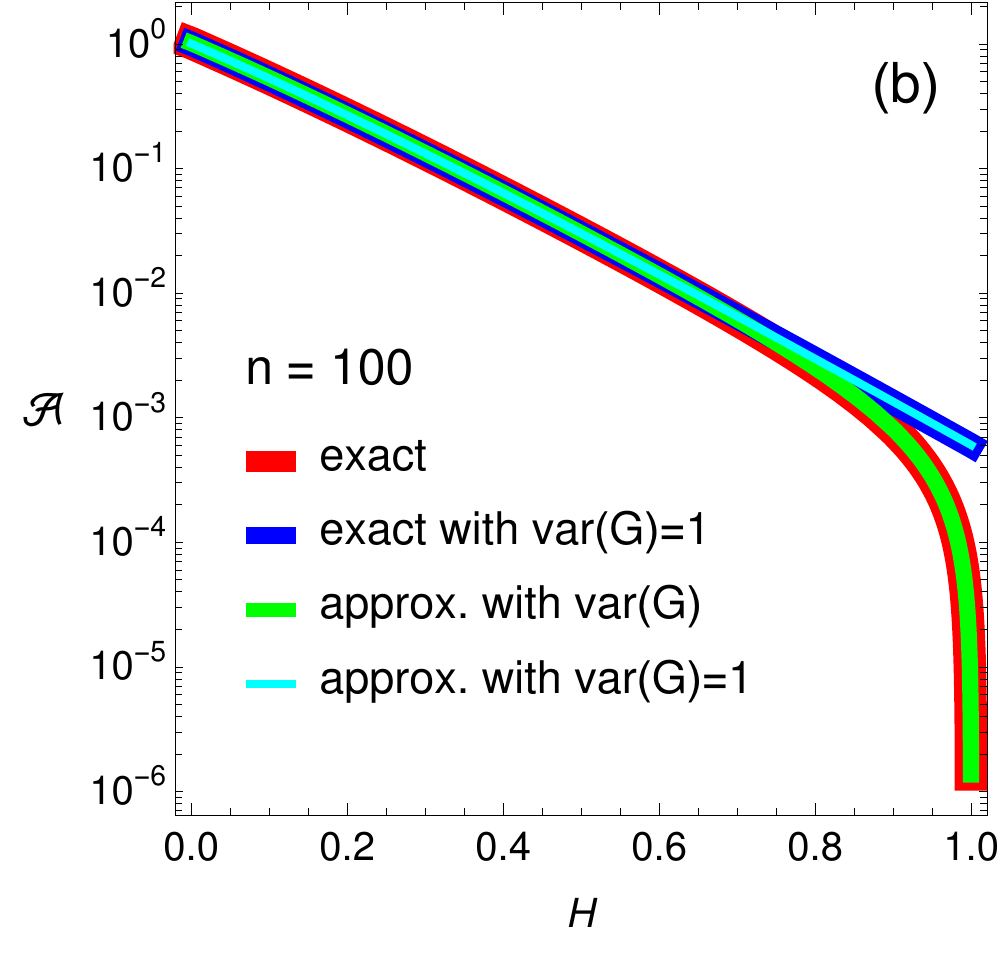}
\includegraphics[width=0.66\columnwidth]{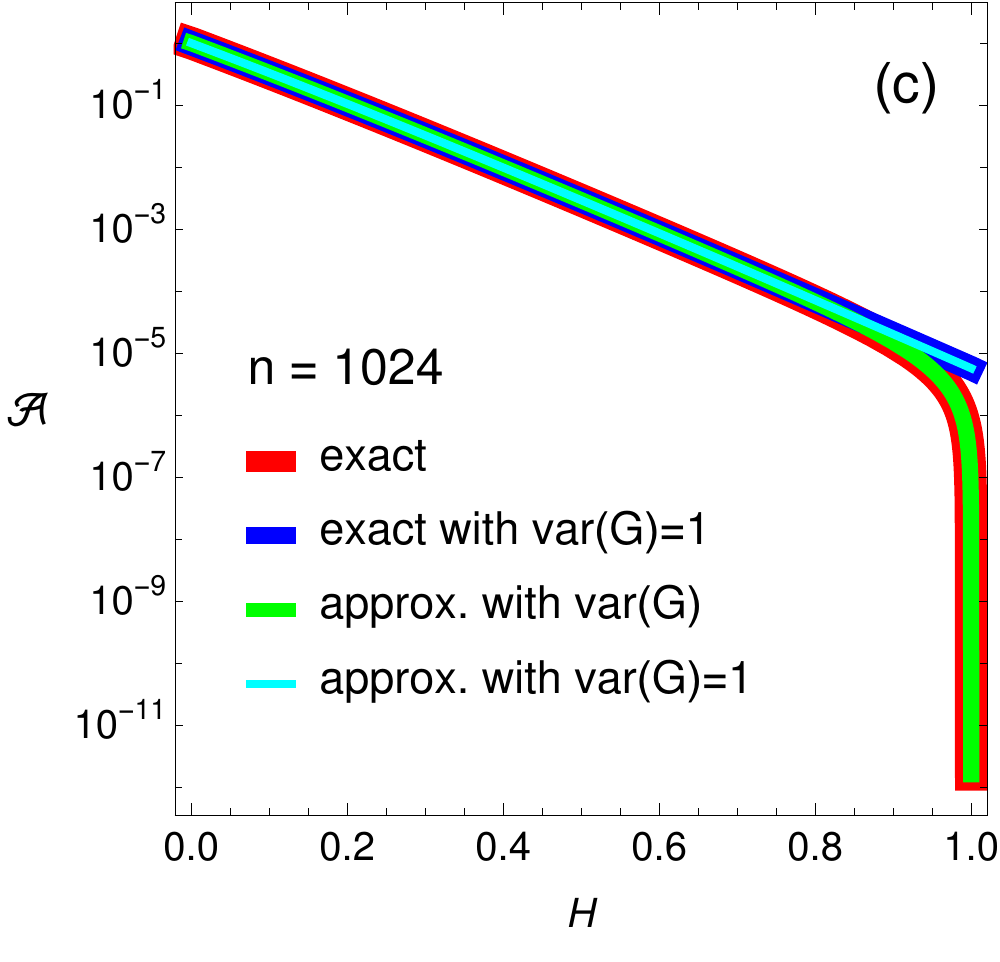}
\caption{$\mathcal{A}(H,n)$ for (a) an extremely short ($n=8$) fBm time series, for (b) a moderate ($n=100$), and (c) a long one ($n=2^{10}$). The ''exact'' line (red) corresponds to Eq.~(\ref{eq7}), ''exact with ${\rm var}(G)=1$'' to the same Eq.~(\ref{eq7}) but with ${\rm var}\left( G_n^H \right)=1$, i.e. set to its asymptotic value (blue), the ''approx. with ${\rm var}(G)$'' denotes Eq.~(\ref{eq19}) with the expression for ${\rm var}\left( G_n^H \right)$ included (green), and ''approx. with ${\rm var}(G)=1$'' is the most simplified form from Eq.~(\ref{eq30}) (cyan). The lines overlap so strongly that for visualization purposes they are depicted with different thickness.}
\label{fig2}
\end{figure*}

These various approximations of $\mathcal{A}_{\rm fBm}(H,n)$ are shown in Fig.~\ref{fig2}. For an unreasonably small $n=8$ one sees discrepancies between the curves for low and moderate $H$, although they are rather small. The finite-size effects, introduced by ${\rm var}\left( G_n^H \right)$ from Eq.~(\ref{eq26}), are significant at higher values of $H$. For a moderate $n=100$, the curves are indistinguishable for most of the range of $H$, with the region of significant influence of ${\rm var}\left( G_n^H \right)$ moved systematically to higher $H$, and for even longer time series (e.g. $n=2^{10}$) the consistency of all curves is only strengthened.  Therefore, the approximation from Eq.~(\ref{eq30}) is a decent one for time series with any reasonable length.


\subsection{Variance of DfGn and Abbe value of fGn}
\label{sect2.2.3}

To obtain an expression for ${\rm var}\left( Y_n^H \right)$ the same methodology from Sect.~\ref{sect2.2.1} is undertaken, i.e. Eq.~(\ref{eq25}) with $G$ changed to $Y$ is employed. Again, developing the sum for a few values of $n$, and utilizng the variance and covariance from Eq.~(\ref{var5}) and Eq.~(\ref{var6}), one can observe the pattern depicted in Table~\ref{tbl1}. Hence one can write
\begin{widetext}
\begin{equation}
{\rm var}\left( Y_n^H \right) = \frac{ 2(2n^2-1) - n^2 2^{2H} + (n-1)^{2H} - 2n^{2H} + (n+1)^{2H} }{n(n-1)}.
\label{eq23}
\end{equation}
\end{widetext}
The exact expression for the Abbe value, taking into account Eq.~(\ref{eq26}), is thence
\begin{widetext}
\begin{equation}
\mathcal{A}_{\rm fGn}(H,n) = \frac{n^{2 H}+(n-2)^{2 H}+n (n-2)\left(4-4^H\right) -2 (n-1)^{2 H}+2-2^{2H}}{2 (n-2) \left(n-n^{2 H-1}\right)}
\label{eq31}
\end{equation}
\end{widetext}

\begin{figure}
\includegraphics[width=0.49\columnwidth]{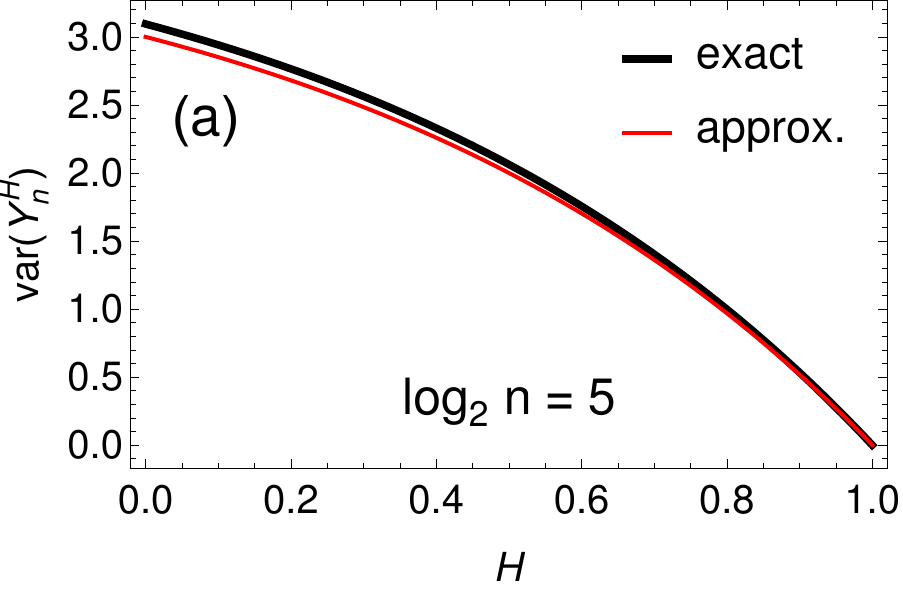}
\includegraphics[width=0.49\columnwidth]{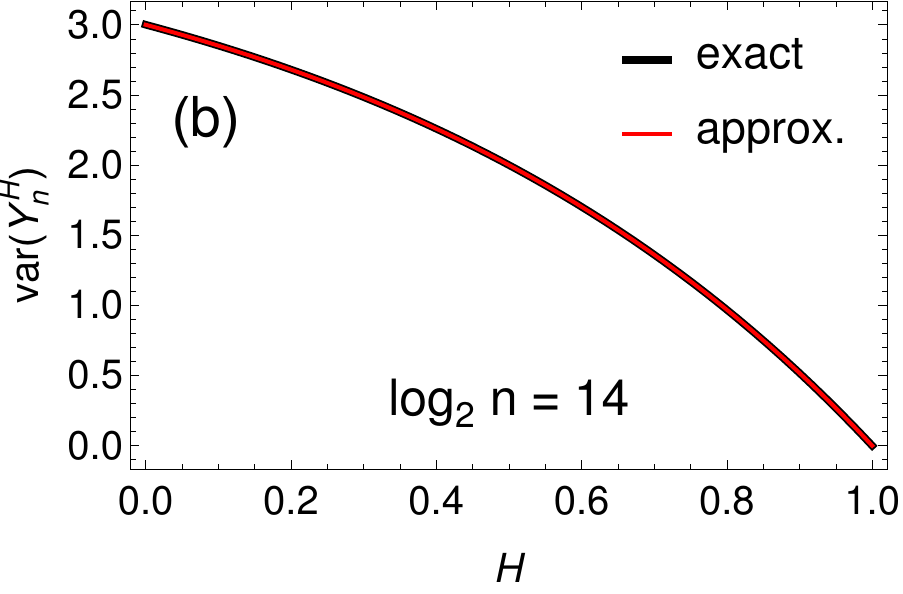}
\caption{${\rm var}\left( Y_n^H \right)$ for (a) a very short ($n=32$) time series, and (b) for a long one, $n=2^{14}$. The ''exact'' line (black) is for Eq.~(\ref{eq23}), and the ''approx.'' line (red) denotes Eq.~(\ref{eq24}).}
\label{fig4}
\end{figure}

In the limit $n\to\infty$, i.e. for any reasonable $n\gg 1$, Eq.~(\ref{eq23}) becomes
\begin{equation}
{\rm var}\left( Y_n^H \right) \approx 4 - 4^H,
\label{eq24}
\end{equation}
i.e. it asymptotically approaches Eq.~(\ref{var5}). Plots of Eq.~(\ref{eq23}) and (\ref{eq24}) are shown in Fig.~\ref{fig4}. For very short time series there is a certain deviation between the expressions, but for higher $n$ the difference is invisible. Therefore, the asymptotic Eq.~(\ref{eq24}) is adequate in any practical scenario.
\begin{table}
\caption{\label{tbl1}Expressions for ${\rm var}\left( Y_n^H \right)$ for first few $n$, and the resulting general formula.}
\begin{ruledtabular}
\begin{tabular}{cccccc}
$n$ & \multicolumn{5}{c}{$n(n-1){\rm var}\left( Y_n^H \right)$} \\
\hline
2  & $2\cdot 7$   & $-2^2\cdot 2^{2H}$  & $+1^{2H}$ & $-2\cdot 2^{2H}$  & $+3^{2H}$  \\
3  & $2\cdot 17$  & $-3^2\cdot 2^{2H}$  & $+2^{2H}$ & $-2\cdot 3^{2H}$  & $+4^{2H}$  \\
4  & $2\cdot 31$  & $-4^2\cdot 2^{2H}$  & $+3^{2H}$ & $-2\cdot 4^{2H}$  & $+5^{2H}$  \\
5  & $2\cdot 49$  & $-5^2\cdot 2^{2H}$  & $+4^{2H}$ & $-2\cdot 5^{2H}$  & $+6^{2H}$  \\
6  & $2\cdot 71$  & $-6^2\cdot 2^{2H}$  & $+5^{2H}$ & $-2\cdot 6^{2H}$  & $+7^{2H}$  \\
7  & $2\cdot 97$  & $-7^2\cdot 2^{2H}$  & $+6^{2H}$ & $-2\cdot 7^{2H}$  & $+8^{2H}$  \\
8  & $2\cdot 127$ & $-8^2\cdot 2^{2H}$  & $+7^{2H}$ & $-2\cdot 8^{2H}$  & $+9^{2H}$  \\
9  & $2\cdot 161$ & $-9^2\cdot 2^{2H}$  & $+8^{2H}$ & $-2\cdot 9^{2H}$  & $+10^{2H}$ \\
10 & $2\cdot 199$ & $-10^2\cdot 2^{2H}$ & $+9^{2H}$ & $-2\cdot 10^{2H}$ & $+11^{2H}$ \\
\hline
   & $2(2n^2-1)$  & $-n^2\cdot 2^{2H} $ & $+(n-1)^{2H}$ & $-2n^{2H}$ & $+(n+1)^{2H}$ \\ 
\end{tabular}
\end{ruledtabular}
\end{table}

The Abbe value is thence
\begin{equation}
\mathcal{A}_{\rm fGn}(H,n) \approx \frac{2-2^{2H-1}}{{\rm var}\left( G_n^H \right)}.
\label{eq28}
\end{equation}
As discussed in Sect.~\ref{sect2.2.1}, the variance of $G_n^H$ can in some instances be approximated by unity. Eq.~(\ref{eq28}) simplifies then to just
\begin{equation}
\mathcal{A}_{\rm fGn}(H,n) \approx 2-2^{2H-1},
\label{eq29}
\end{equation}
independent on the length $n$ of the time series.

The asymptotic Eq.~(\ref{eq29}) ranges from $0$ to $3/2$ when $H$ decreases from 1 to 0. The expression from Eq.~(\ref{eq28}) reaches its maximum of $\frac{3}{2}\frac{n}{n+1}$ at $H=0$. The asymptotic minimum, as $H\to 1$, is $\frac{n-1}{n}\frac{\ln 4}{\ln n} = \frac{n-1}{n}\log_n4$. For $n=2^{14}$, these values are 1.49991 and 0.14285, respectively, in perfect agreement with Fig.~3 in \cite{tarnopolski16}.

\section{Representation of fBm and fGn in the $\mathcal{A}-\mathcal{T}$ plane}
\label{sect2.3}

\begin{figure*}
\includegraphics[width=0.8\columnwidth]{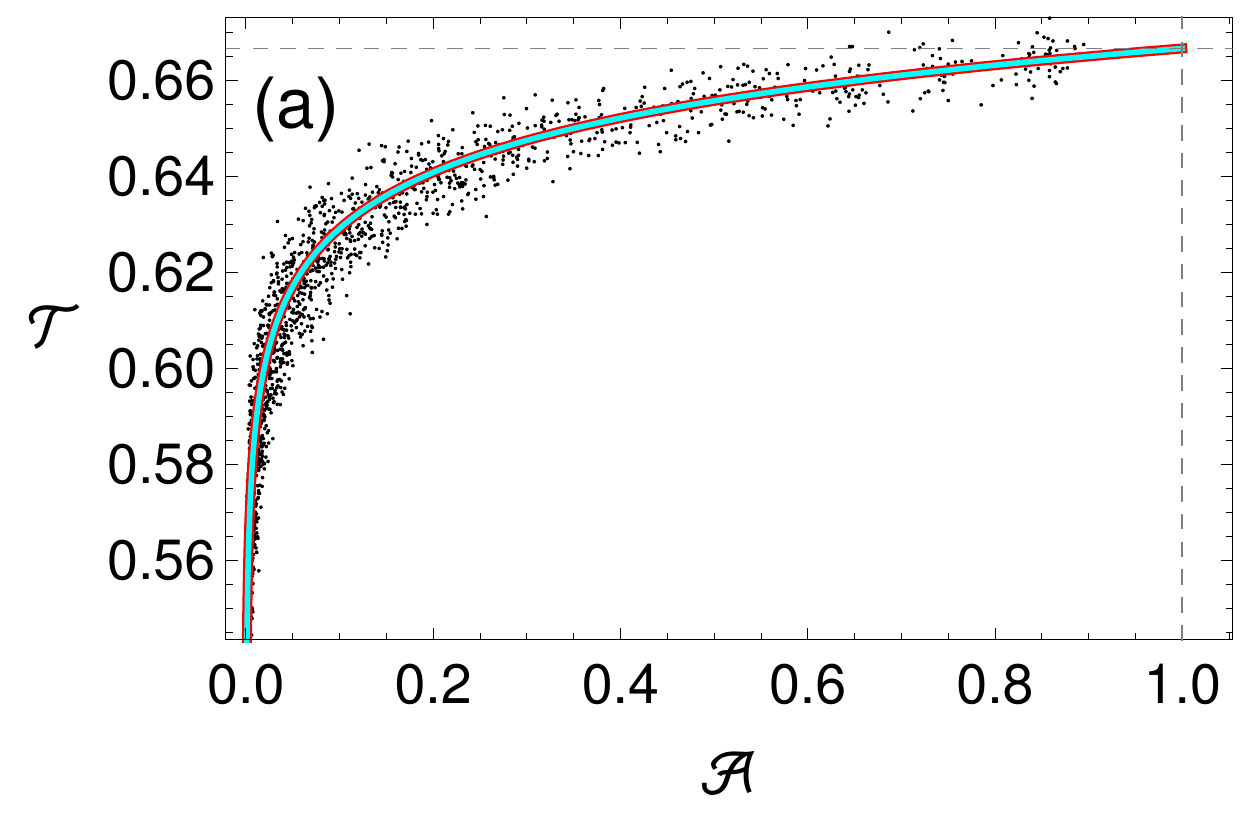} 
\includegraphics[width=0.8\columnwidth]{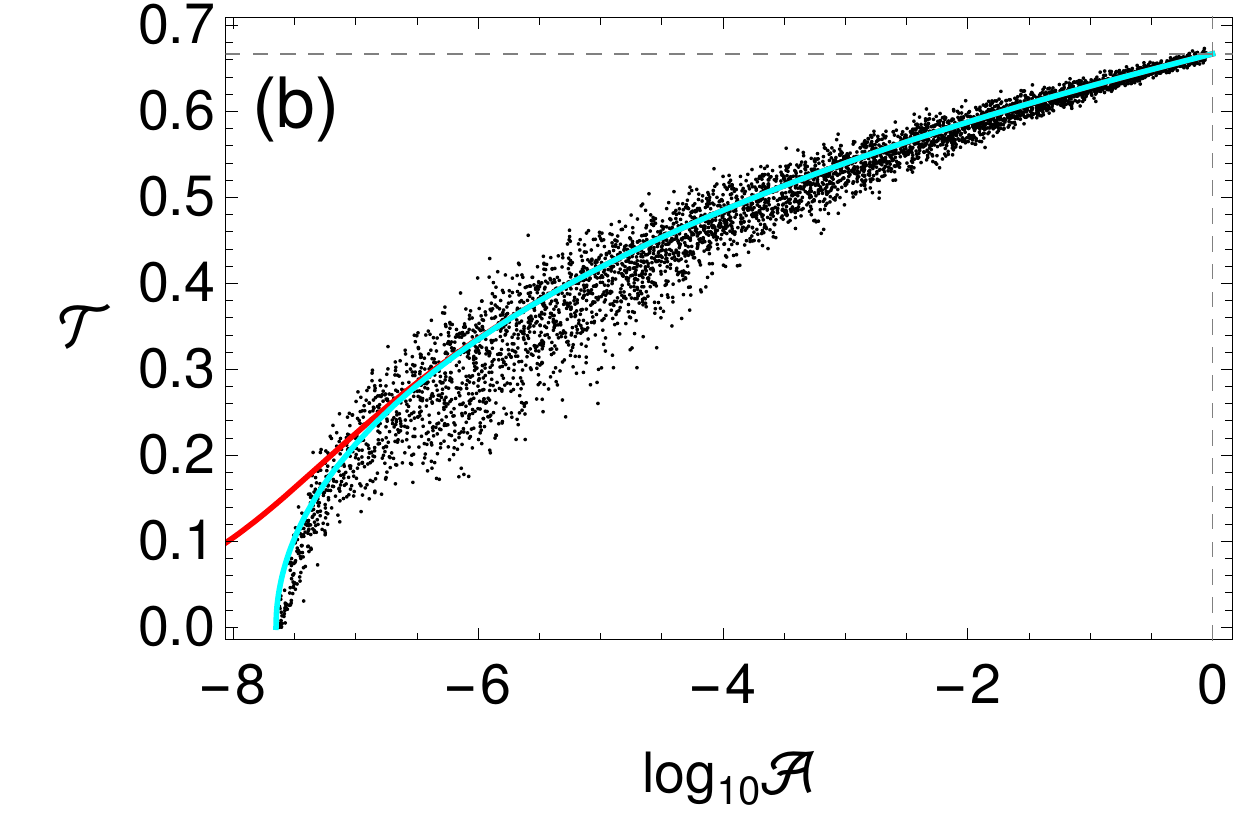} \\
\includegraphics[width=0.8\columnwidth]{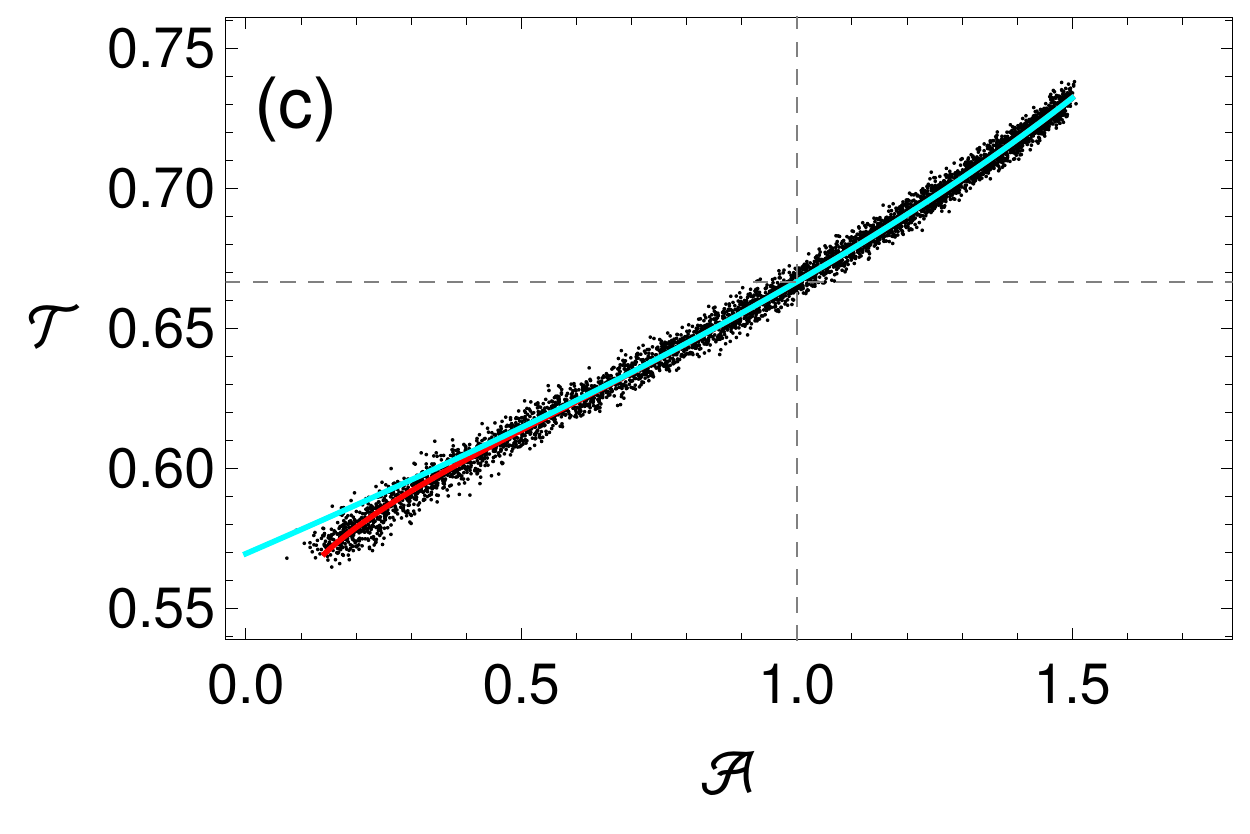}
\includegraphics[width=0.8\columnwidth]{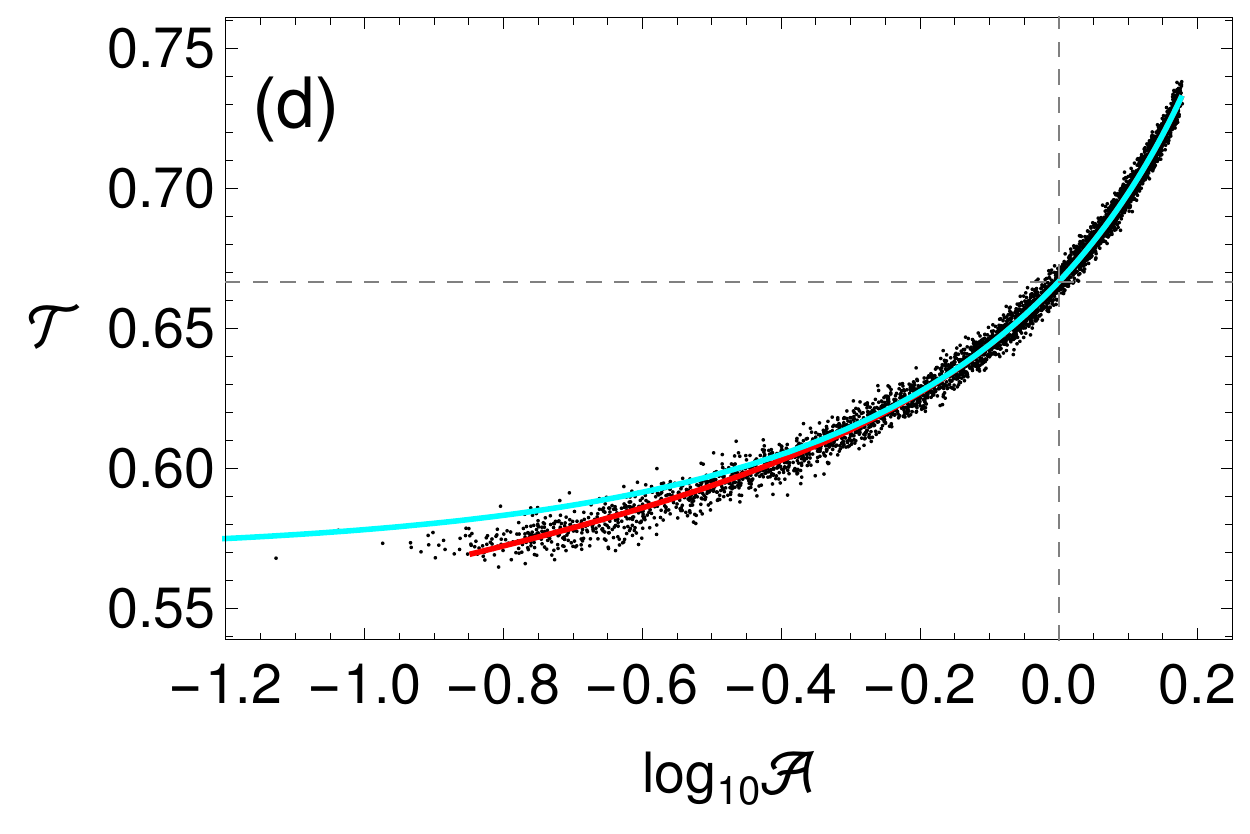}
\caption{The $\mathcal{A}-\mathcal{T}$ plane for $n=2^{14}$: (a)--(b) fBm, (c)--(d) fGn. The red lines utilize the exact expressions for $\mathcal{A}$ [i.e. Eq.~(\ref{eq7}) and (\ref{eq31})], and cyan ones use the approximations [Eq.~(\ref{eq30}) for fBm and the asymptotic Eq.~(\ref{eq29}) for fGn]. Eq.~(\ref{eq1}) and (\ref{eq32}) describe $\mathcal{T}_{\rm fBm}$ and $\mathcal{T}_{\rm fGn}$, respectively. The horizontal and vertical dashed lines mark $\mathcal{T}=2/3$ and $\mathcal{A}=1$, respectively. Panels (b) and (d) display the same as (a) and (c), but with a logarithmic horizontal axis. The discrepancies in (b) at very low $\mathcal{A}$ are due to the deviation of ${\rm var}\left( G_n^H \right)$ from unity when $H$ tends to 1. }
\label{fig3}
\end{figure*}

The $\mathcal{A}-\mathcal{T}$ plane is displayed in Fig.~\ref{fig3}. The black points come from simulations \cite{tarnopolski16}. In case of fBm [Fig.~\ref{fig3} (a) and (b)], the red line employs the exact formula for $\mathcal{A}_{\rm fBm}$ from Eq.~(\ref{eq7}), while the cyan line depicts the approximation from Eq.~(\ref{eq30}). Eq.~(\ref{eq1}) describes $\mathcal{T}_{\rm fBm}$.

In case of fGn [Fig.~\ref{fig3} (c) and (d)], the red line corresponds to the exact Eq.~(\ref{eq31}) for $\mathcal{A}_{\rm fGn}$, while the cyan line to the asymptotic form from Eq.~(\ref{eq29}). Eq.~(\ref{eq32}) describes $\mathcal{T}_{\rm fGn}$. Panels (b) and (d) of Fig.~\ref{fig3} employ a logarithmic horizontal axis to fully display the dependence $\mathcal{T}(\mathcal{A})$ at small values of $\mathcal{A}$ (i.e. high values of $H$). The agreement between numerical simulations and the analytic description is very good; also the approximations for $\mathcal{A}_{\rm fGn}(H,n)$ work well in the $\mathcal{A}-\mathcal{T}$ plane. In particular, the approximation from Eq.~(\ref{eq28}) for fGn is as good as the exact Eq.~(\ref{eq31}).

\section{ARMA processes}
\label{sect2.4}

The methodology from Sect.~\ref{sect2.1} and \ref{sect2.2} is applicable also to ARMA processes. Generalizing Eq.~(\ref{eq2}), similarly as was done in Eq.~(\ref{eq3}) and (\ref{eq21}), the Abbe value of a process $X$ is
\begin{equation}
\mathcal{A} = \frac{1}{2} \frac{{\rm var}\left( dX \right)}{{\rm var}\left( X \right)},
\label{eq38true}
\end{equation}
where $dX$ denotes the increments of $X$. The fraction of turning points is given by Eq.~(\ref{eq36})--(\ref{eq38}). It will be convenient to express $\rho(d)$ in terms of the autocorrelation function $\rho_d=E\left[ X(t)X(t+d) \right]$:
\begin{equation}
\rho(d) = \frac{2\rho_d-1-\rho_{2d}}{2(1-\rho_d)},
\label{}
\end{equation}
so that Eq.~(\ref{eq36}) becomes
\begin{equation}
p_{123}(d) = \frac{1}{\pi}\arcsin\left( \frac{1}{2}\sqrt{\frac{1-\rho_{2d}}{1-\rho_d}} \right).
\label{eq40}
\end{equation} 
This formula can be directly applied also to fGn and DfGn, but not to fBm which is nonstationary.

For ARMA processes, $\rho_d$ and ${\rm var}\left( X \right)$ are easily obtainable \cite{brockwell,brockwell2}. The variance of the differentiated process, ${\rm var}\left( dX \right)$, is calculated as
\begin{equation}
\begin{split}
{\rm var}\left( dX(t) \right) =& E\left[ dX^2(t) \right] \\
=& E\left[ \left( X(t+1)-X(t) \right)^2 \right] \\
=& E\left[ X^2(t+1) \right] + E\left[ X^2(t) \right] \\
 &- 2E\left[ X(t+1)X(t) \right].
\end{split}
\label{eq41}
\end{equation}

\subsection{AR(1)}
\label{AR1}

Consider a weakly stationary process $X_t = a_1 X_{t-1} + \varepsilon_t$, $-1<a_1<1$, where $\varepsilon_t$ is a white-noise error term. Then
\begin{equation}
\rho_d = a_1^d
\label{}
\end{equation}
for all $d$, thus
\begin{equation}
\mathcal{T} = 1-\frac{2}{\pi}\arcsin\left( \frac{1}{2}\sqrt{1+a_1} \right),
\label{}
\end{equation}
reaching its minimum of $1/2$ when $a_1\rightarrow 1$, and its maximum of $1$ when $a_1\rightarrow -1$. 

One then obtains
\begin{equation}
{\rm var}\left( X \right) = \frac{1}{1 - a_1^2}
\label{}
\end{equation}
and
\begin{equation}
{\rm var}\left( dX \right) = \frac{2}{1+a_1},
\label{}
\end{equation}
and thus
\begin{equation}
\mathcal{A} = 1-a_1,
\label{}
\end{equation}
reaching its minimum of $0$ when $a_1\rightarrow 1$, and maximum of $2$ when $a_1\rightarrow -1$. One can then write $\mathcal{T}$ explicitly as a function of $\mathcal{A}$: 
\begin{equation}
\mathcal{T} = 1-\frac{2}{\pi}\arcsin\left( \frac{1}{2}\sqrt{2-\mathcal{A}} \right),
\label{eq47}
\end{equation}
which is displayed in Fig.~\ref{fig6}. The Ornstein-Uhlenbeck (OU) process, a continuous analog of AR(1), yields the same Eq.~(\ref{eq47}), but restricted to $\mathcal{A}\in[0,1]$ (see Appendix~\ref{app}).

\subsection{AR(2)}
\label{AR2}

Consider a weakly stationary process $X_t = a_1 X_{t-1} + +a_2 X_{t-2} + \varepsilon_t$, $-1<a_2<1 \wedge a_2<1+a_1 \wedge a_2<1-a_1 $. Then $\rho_d$ is given by a recurrent relation (the Yule-Walker equations):
\begin{equation}
\rho_d = 
\begin{cases}
 1 & d=0 \\
 \frac{a_1}{1-a_2} & d = 1 \\
 a_1\rho_{d-1} + a_2\rho_{d-2} & d\geq 2
\end{cases}
\label{}
\end{equation}
thus
\begin{equation}
\mathcal{T} = 1-\frac{2}{\pi}\arcsin\left( \frac{1}{2}\sqrt{1+a_1-a_2} \right),
\label{eq55}
\end{equation}
reaching its minimum of $0$ when $a_1\rightarrow 2, a_2\rightarrow -1$, and its maximum of $1$ along the line $a_2=a_1+1$.

One then obtains
\begin{equation}
{\rm var}\left( X \right) = \frac{1-a_2}{(a_2+1) (a_2-a_1-1) (a_2+a_1-1)}
\label{}
\end{equation}
and
\begin{equation}
{\rm var}\left( dX \right) = \frac{2}{1+a_1+a_1 a_2-a_2^2},
\label{}
\end{equation}
thus
\begin{equation}
\mathcal{A} = 1+\frac{a_1}{a_2-1},
\label{eq58}
\end{equation}
reaching its minimum of $0$ along the line $a_2 = 1-a_1$, and maximum of $2$ along the line $a_2 = 1+a_1$. One cannot write $\mathcal{T}$ explicitly as a function of $\mathcal{A}$, as $(\mathcal{A},\mathcal{T})$ is a two-dimensional region, depicted in Fig.~\ref{fig6}. However, it is possible to give a simple formula for the boundaries of this region. Note that for a given $a_1$, $\mathcal{T}$ is minimal when $a_2\rightarrow -1$. Hence by setting $a_2=-1$, one can then solve Eq.~(\ref{eq58}) for $a_1$, i.e. write $a_1=2(1-\mathcal{A})$, and insert this into Eq.~(\ref{eq55}) to obtain the lower boundary as
\begin{equation}
\mathcal{T}\left( \mathcal{A} \right)_{\rm lower\, boundary} = \frac{2}{\pi}\arcsin\left( \sqrt{\frac{\mathcal{A}}{2}} \right).
\label{eq52}
\end{equation}
The upper boundary, $\mathcal{T}=1$, is attained when $a_2\rightarrow 1$, and from the left the region is a vertical line $\mathcal{A}=0$, obtained by sweeping $a_2$ from $-1$ to $1$.

Notice that AR(1) is a special case of AR(2) with $a_2=0$. One can then observe, by repeating the above reasoning for an arbitrary $a_2$, that the region of availability, $\mathcal{S}_{{\rm AR}(2)}$, is formed as a continuum of curves parametrized by $a_2$:
\begin{equation}
\begin{split}
& \mathcal{S}_{{\rm AR}(2)} = \left\{ \mathcal{T}\left( \mathcal{A} \right) \big|a_2 \right\} \\
& = \left\{ 1-\frac{2}{\pi}\arcsin\left( \frac{1}{2}\sqrt{(\mathcal{A}-2)(a_2-1)} \right) \Bigg| -1 < a_2 < 1 \right\},
\end{split}
\label{}
\end{equation}
reducing to Eq.~(\ref{eq52}) when $a_2\to -1$. 

\subsection{MA(1)}

Consider $X_t = b_1\varepsilon_{t-1} + \varepsilon_t$, weakly stationary for all $b_1\in\mathbb{R}$. The autocorrelation function is
\begin{equation}
\rho_d = 
\begin{cases}
 1 & d=0 \\
 \frac{b_1}{1+b_1^2} & d = 1 \\
 0 & {\rm otherwise}
\end{cases}
\label{}
\end{equation}
thus
\begin{equation}
\mathcal{T} = 1-\frac{2}{\pi}\arcsin\left( \frac{1}{2}\sqrt{\frac{1+b_1^2}{1-b_1+b_1^2}} \right),
\label{}
\end{equation}
reaching its minimum of $1/2$ at $b_1=1$, and its maximum of $\left(2/\pi\right)\arcsec\sqrt{6}\approx 0.73$ at $b_1=-1$. 

One then obtains
\begin{equation}
{\rm var}\left( X \right) = 1 + b_1^2
\label{}
\end{equation}
and
\begin{equation}
{\rm var}\left( dX \right) = 2\left( 1+b_1(b_1-1) \right),
\label{}
\end{equation}
thus
\begin{equation}
\mathcal{A} = \frac{1-b_1+b_1^2}{1+b_1^2},
\label{}
\end{equation}
reaching its minimum of $1/2$ at $b_1=1$, and maximum of $3/2$ at $b_1=-1$. One can then write $\mathcal{T}$ explicitly as a function of $\mathcal{A}$: 
\begin{equation}
\mathcal{T} = 1-\frac{2}{\pi}\arcsin\left( \frac{1}{2\sqrt{\mathcal{A}}} \right),
\label{}
\end{equation}
which is displayed in Fig.~\ref{fig6}.

\subsection{MA(2)}

Consider $X_t = b_1\varepsilon_{t-1} + b_2\varepsilon_{t-2} + \varepsilon_t$, weakly stationary for all $b_1,b_2\in\mathbb{R}$, for which
\begin{equation}
\rho_d = 
\begin{cases}
 1 & d=0 \\
 \frac{b_1(1+b_2)}{1+b_1^2+b_2^2} & d = 1 \\
 \frac{b_2}{1+b_1^2+b_2^2} & d = 2 \\
 0 & {\rm otherwise}
\end{cases}
\label{}
\end{equation}
thus
\begin{equation}
\mathcal{T} = 1-\frac{2}{\pi}\arcsin\left( \frac{1}{2}\sqrt{\frac{1+b_1^2+b_2(b_2-1)}{1+b_1^2+b_2^2-b_1(b_2+1)}} \right),
\label{}
\end{equation}
reaching its minimum of $2/5$ at $b_1=\frac{1}{2}\left( 1+\sqrt{5} \right), b_2=1$, and its maximum of $4/5$ at $b_1=\frac{1}{2}\left( 1-\sqrt{5} \right), b_2=1$.

One then obtains
\begin{equation}
{\rm var}\left( X \right) = 1 + b_1^2 + b_2^2
\label{}
\end{equation}
and
\begin{equation}
{\rm var}\left( dX \right) = 2\left( 1+b_1^2+b_2^2-b_1(b_2+1) \right),
\label{}
\end{equation}
thus
\begin{equation}
\mathcal{A} = 1-\frac{b_1(b_2+1)}{1+b_1^2+b_2^2},
\label{}
\end{equation}
reaching its minimum of $1-1/\sqrt{2}$ at $b_1=\sqrt{2},b_2=1$, and maximum of $1+1/\sqrt{2}$ at $b_1=-\sqrt{2},b_2=1$. One cannot write $\mathcal{T}$ explicitly as a function of $\mathcal{A}$, as $(\mathcal{A},\mathcal{T})$ is a two-dimensional region, depicted in Fig.~\ref{fig6}. 

\subsection{ARMA(1,1)}

Consider $X_t = a_1 X_{t-1} + b_1\varepsilon_{t-1}+ \varepsilon_t$, weakly stationary for $-1<a_1<1$ and all $b_1\in\mathbb{R}$, for which
\begin{equation}
\rho_d = 
\begin{cases}
 1 & d=0 \\
 \frac{a_1^{d-1} (a_1+b_1) (a_1 b_1 + 1)}{1 + 2 a_1 b_1 + b_1^2} & {\rm otherwise}
\end{cases}
\label{}
\end{equation}
thus
\begin{equation}
\mathcal{T} = 1-\frac{2}{\pi}\arcsin\left( \frac{1}{2}\sqrt{\frac{(1+a_1)(1+a_1 b_1+b_1^2)}{1+b_1(a_1+b_1-1)}} \right),
\label{eq67}
\end{equation}
reaching its minimum of $1/3$ when $a_1\rightarrow 1, b_1 = 1$, and its maximum of $1$ when $a_1\rightarrow -1$, along the line $b_1\in\mathbb{R}$.

\begin{figure}
\raggedleft
\includegraphics[width=\columnwidth]{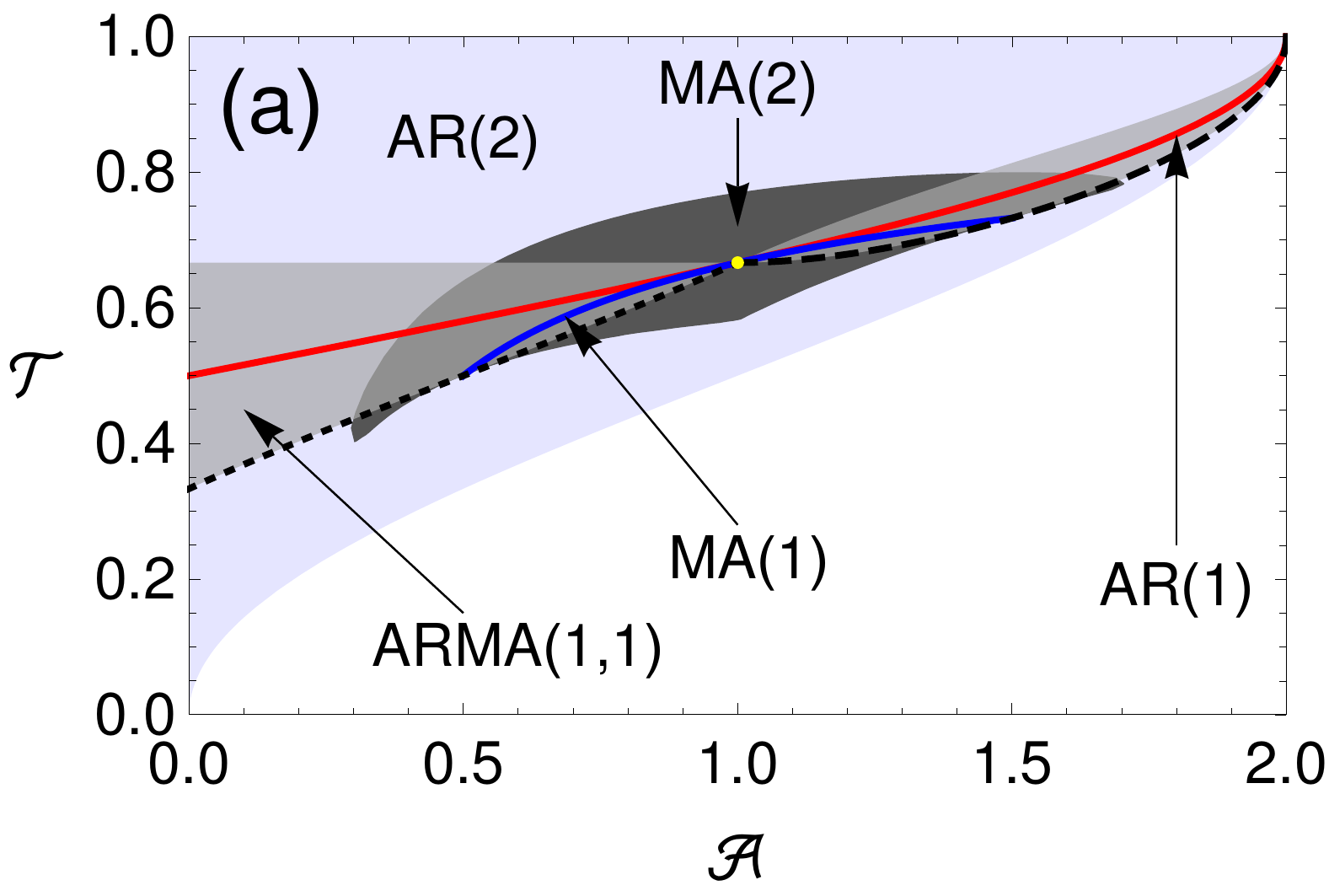}
\includegraphics[width=0.939\columnwidth]{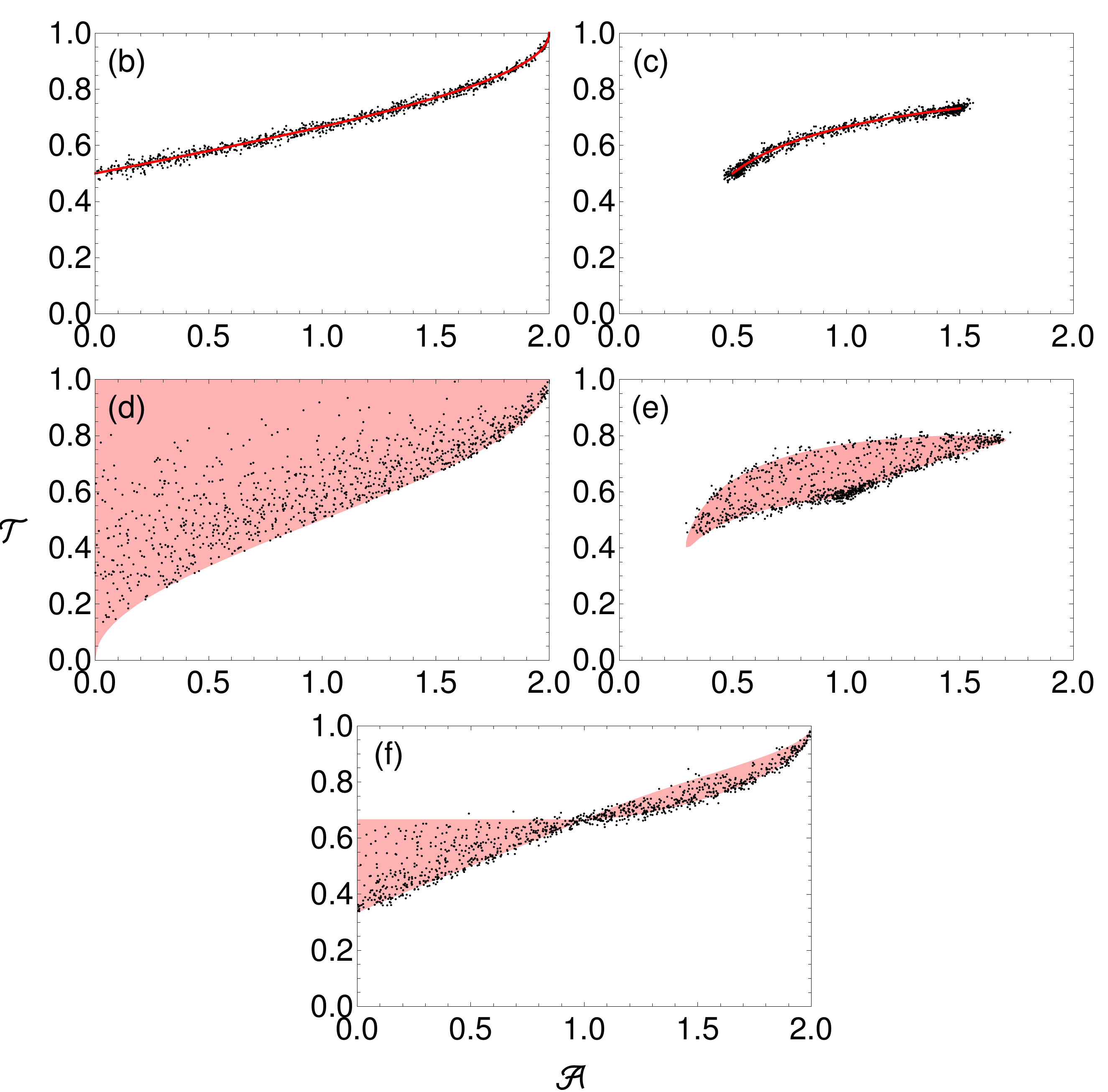}
\caption{(a) Regions of the $\mathcal{A}-\mathcal{T}$ plane available for ARMA processes: AR(1) (red line), MA(1) (blue line), MA(2) (darker gray region), AR(2) (light blue in the background), and ARMA(1,1) (lighter gray region). The black dotted line in the range $\mathcal{A}\in (0,1)$ is Eq.~(\ref{eq_it_2}) for ARMA(1,1) with $b_1=1$, and the black dashed line in $\mathcal{A}\in (1,2)$ is Eq.~(\ref{eq_it_1}) for ARMA(1,1) with $b_1=-1$. These two lines mark the lower boundary of the region of availability for the ARMA(1,1) process. The yellow dot at $(1,2/3)$ denotes white noise. In the other panels, these regions are shown together with locations of $10^3$ simulated processes: (b) AR(1), (c) MA(1), (d) AR(2), (e) MA(2), and (f) ARMA(1,1). In the simulations, the AR coefficients were drawn uniformly from the respective regions fulfilling the stationarity conditions, and the MA coefficients were drawn uniformly from $(-1,1)$. Note the varying density of points in the regions of availability. }
\label{fig6}
\end{figure}

One then obtains
\begin{equation}
{\rm var}\left( X \right) = \frac{1 + 2 a_1 b_1 + b_1^2}{1-a_1^2}
\label{}
\end{equation}
and
\begin{equation}
{\rm var}\left( dX \right) = 2\frac{1+b_1 (a_1+b_1-1)}{a_1+1},
\label{}
\end{equation}
thus
\begin{equation}
\mathcal{A} = 1 - a_1 + \frac{b_1(a_1^2-1)}{1+2a_1 b_1+b_1^2},
\label{eq70}
\end{equation}
reaching its minimum of $0$ when $a_1\rightarrow 1$, along the line $b_1\in\mathbb{R}$, and maximum of $2$ when $a_1\rightarrow -1$, along the line $b_1\in\mathbb{R}$. One cannot write $\mathcal{T}$ explicitly as a function of $\mathcal{A}$, as $(\mathcal{A},\mathcal{T})$ is a two-dimensional region, depicted in Fig.~\ref{fig6}.

Note that Eq.~(\ref{eq67}) and (\ref{eq70}) are invariant on changing $b_1$ to $1/b_1$, so that in the context of geometrical depiction of the region of availability only the range $-1\leq b_1\leq 1$ needs to be considered. Roughly speaking, cases with $|b_1|\gg 1$ are equivalent to $|b_1|\ll 1$. However, $b_1=\pm 1$ are special instances:
\begin{itemize}
\item when $b_1=-1$, Eq.~(\ref{eq70}) yields $a_1=3-2\mathcal{A}$, which fulfills the stationarity condition, $a_1\in (-1,1)$, only for $\mathcal{A}\in (1,2)$. In this range:
\begin{equation}
\mathcal{T}\left({b_1=-1}\right) = 1-\frac{2}{\pi}\arcsin\left( \frac{1}{2}\sqrt{5-\frac{2}{\mathcal{A}}-2\mathcal{A}} \right);
\label{eq_it_1}
\end{equation}
\item likewise, when $b_1=1$, one obtains $a_1=1-2\mathcal{A}$, which leads to $\mathcal{A}\in (0,1)$, giving:
\begin{equation}
\mathcal{T}\left({b_1=1}\right) = 1-\frac{2}{\pi}\arcsin\left( \frac{1}{2}\sqrt{3-2\mathcal{A}} \right).
\label{eq_it_2}
\end{equation}
\end{itemize}
In particular, $b_1=0$ reduces an ARMA(1,1) process to an AR(1) one, reproducing respective formulae from Sect.~\ref{AR1}.

Similarly as was done in Sect.~\ref{AR2} for AR(2) processes, the region of availabity for ARMA(1,1) can be described as a continuum of curves, $\mathcal{S}_{\rm ARMA(1,1)} = \left\{ \mathcal{T} \left( \mathcal{A} \right) \big| b_1 \right\}$, parametrized by $b_1$: one needs to solve Eq.~(\ref{eq70}) for $a_1$ and insert the solution into Eq.~(\ref{eq67}). The resulting formula, easy to derive but of a quite complicated and noninformative form, is not displayed herein.

\section{Applications}
\label{sect::applic}

\subsection{Bacterial cytoplasm}
\label{bacteria}

\begin{figure}
\centering
\includegraphics[width=\columnwidth]{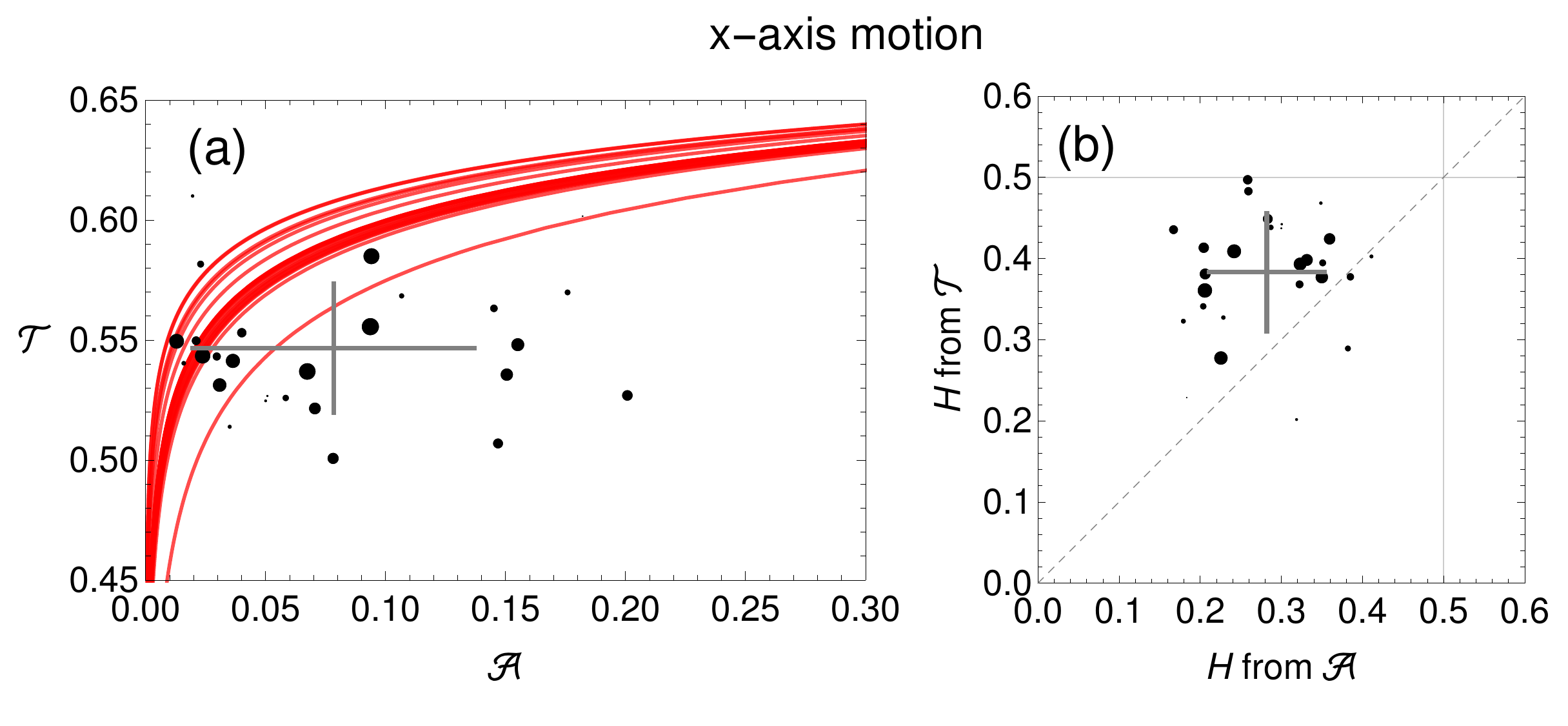}
\caption{(a) Locations in the $\mathcal{A}-\mathcal{T}$ plane of the 27 tracks of $x$-axis motion of mRNA molecules inside {\it E. coli}. The red lines are the fBm lines, corresponding to the length of time series $140\leq n\leq 1628$. Lower curves correspond to lower $n$. (b) Estimated $H$ values, obtained from the formulae for $\mathcal{A}$ and $\mathcal{T}$, i.e. Eq.~(\ref{eq30}) and (\ref{eq1}), respectively. In both panels, the size of the point is proportional to $n$. The gray crosses symbolize the (unweighted) means and standard deviations of the displayed locations.}
\label{fig9}
\end{figure}

The two-dimensional motion $\left[ x(t), y(t) \right]$ of individual mRNA molecules inside live {\it Escherichia coli} bacteria were tracked in \citep{golding06}. It was found that they follow anomalous diffusion, with $H<0.5$, confirmed by other methods as well \citep{magdziarz09}. Herein, the time series $x$ and $y$ are treated separately. Results for the 27 tracks are displayed in Fig.~\ref{fig9} for the $x$-axis. Similar outcomes were obtained for the $y$-axis. The locations in the $\mathcal{A}-\mathcal{T}$ plane are in agreement with an fBm description, and the values extracted using Eq.~(\ref{eq1}) and (\ref{eq30}), yielding $0.2\lesssim H\lesssim 0.5$, are consistent with each other, and confirm that the observed process is indeed subdiffusive.

\subsection{Zeros of the Riemann $\zeta$}
\label{zeros}

\begin{figure}
\centering
\includegraphics[width=0.7\columnwidth]{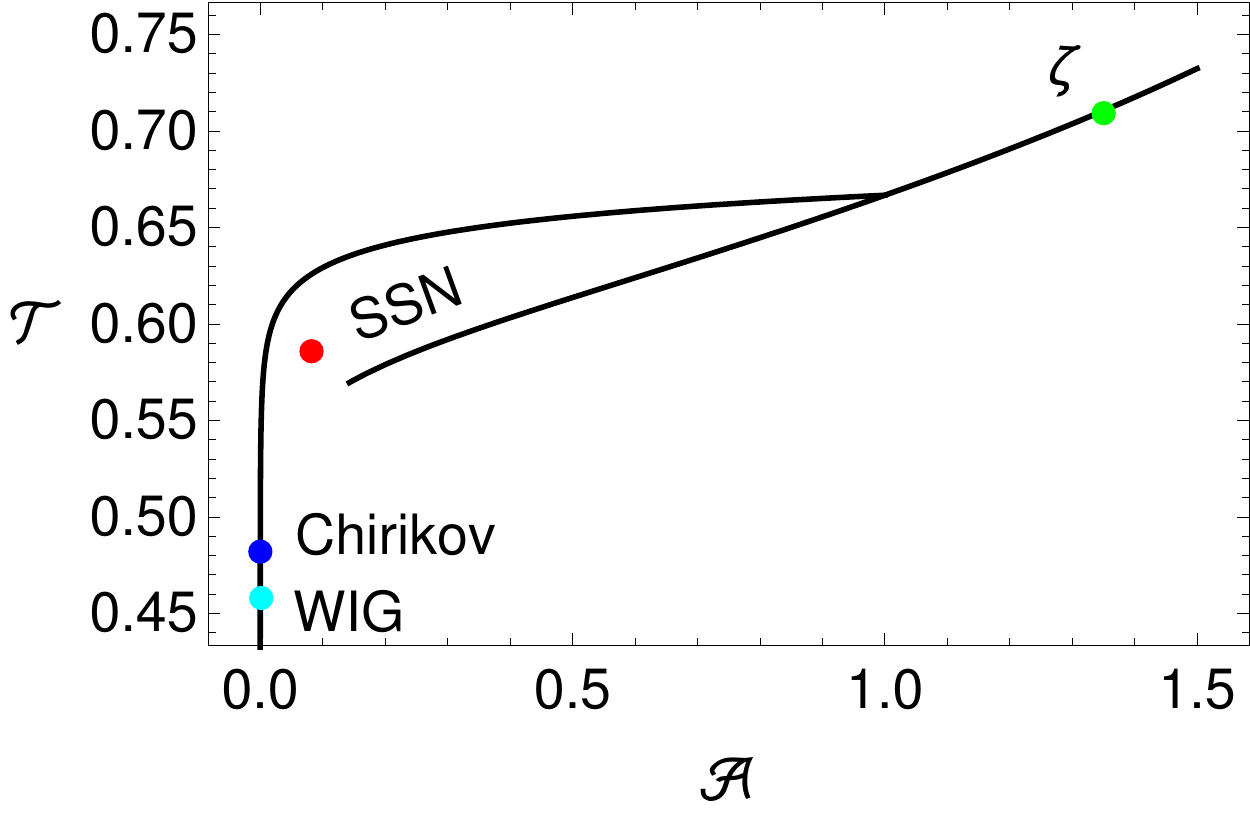}
\caption{Locations of some real-world and generated time series. The black lines correspond to fBm and fGn with $n=2^{14}$. See text for details. }
\label{applications}
\end{figure}

The first $10^6+1$ nontrivial zeros, $\frac{1}{2}+{\rm i}\gamma_n$, of the Riemann $\zeta$ function \citep{platt15} were retrieved \footnote{\url{https://www.lmfdb.org/zeros/zeta/}}. Normalized spacings between consecutive zeros were computed as \citep{odlyzko87}
\begin{equation}
\delta_n = \frac{ \gamma_{n+1} - \gamma_n }{2\pi} \ln\left( \frac{\gamma_n}{2\pi} \right).
\label{}
\end{equation}
The location of this sequence in the $\mathcal{A}-\mathcal{T}$ plane is $(1.350,0.709)$. This is remarkably close to the fGn line (Fig.~\ref{applications}), and Eq.~(\ref{eq32}) and (\ref{eq29}) give $H=0.19$. In comparison, the discrete wavelet transform (DWT) method \citep{tarnopolski16} returns $H=0.06$, hence also strongly implying $H<0.5$. However, as the distribution of $\delta_n$ is not normal, but rather follows the distribution of the Gaussian unitary ensemble (GUE) according to the GUE hypothesis \citep{montgomery73}, so this sequence is not a Gaussian process, strictly speaking. Note, however, that a location in the $\mathcal{A}-\mathcal{T}$ plane can be computed for any type of data; in this case, due to $\mathcal{A}>1$ and $\mathcal{T}>2/3$, one gets a clear information that the series is---in a sense---(much) more noisy than regular white noise.

\subsection{Sunspot numbers}
\label{ssn}

The curently available from the World Data Center Sunspot Index and Long-term Solar Observations \footnote{SILSO World Data Center, Royal Observatory of Belgium, Brussels, \url{http://www.sidc.be/silso}} sample of 3250 monthly SNN are described by $H=0.3$, as computed with the DWT approach \footnote{Similar value can be obtained with data in \citep{tarnopolski16} when the two outliers on the left of Fig.~6b therein are discarded.}. The location in the $\mathcal{A}-\mathcal{T}$ plane is $(0.082, 0.586)$, Fig.~\ref{applications}, and Eq.~(\ref{eq1}) and (\ref{eq30}) give $H\approx 0.18-0.27$. This is in agreement with some works \citep{movahed06} that also compute $H<0.5$. Note that the SSN sequence is slightly off the fBm line, hence it not necessarily need to be adequately modeled by an fBm process. The SSN is a straightforward way of monitoring the Sun's activity, and since the sunspots are tightly connected with the magnetic fields governing solar flares and coronal mass ejections, its proper modeling is crucial in forecasting the space weather conditions.

\subsection{Chaos}
\label{chaos}

The Chirikov standard map for a chaotic state (in an unbounded setting) was examined in \citep{tarnopolski16}. Its location in the $\mathcal{A}-\mathcal{T}$ plane is $(0.0002, 0.482)$, and lies directly on the fBm line, Fig.~\ref{applications}, and Eq.~(\ref{eq1}) and (\ref{eq30}) give $H\approx 0.5$, which is in perfect agreement with the estimate with DWT method, yielding $H=0.48$. This means that, in the context of long-term memory, it is uncorrelated, and acts like Brownian motion.

\begin{figure}
\centering
\includegraphics[width=\columnwidth]{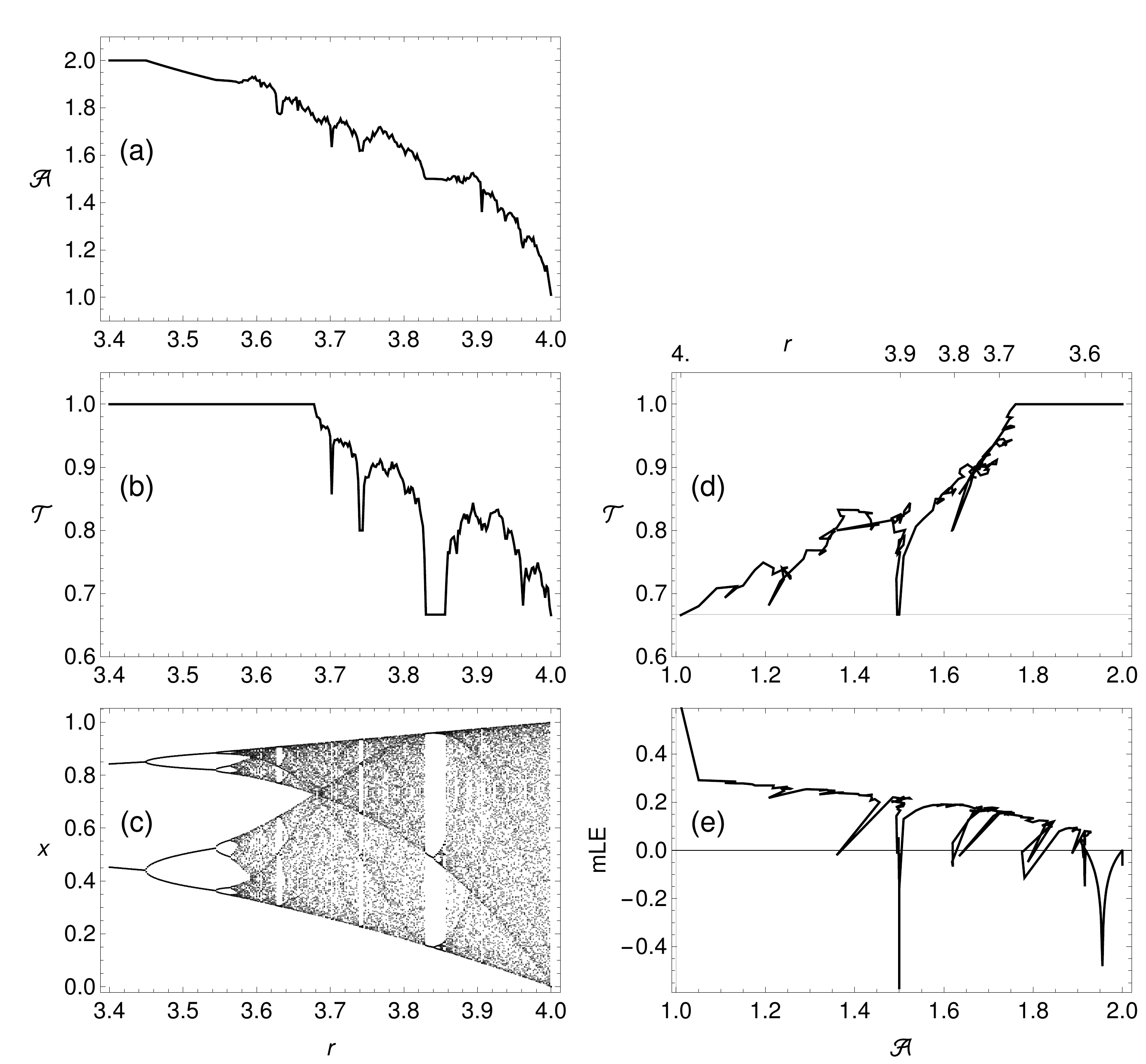}
\caption{Relations between $\mathcal{A}$, $\mathcal{T}$, $r$, and mLE for the logistic map.}
\label{logistic}
\end{figure}

To further illustrate chaotic behavior in the $\mathcal{A}-\mathcal{T}$ plane, the logistic map $x_{i+1} = r x_i (1 - x_i)$ is considered. For $r\in[3.4, 4.0]$, with a step $\Delta r = 0.002$, time series of length $n=10^4$ were generated and their $(\mathcal{A},\mathcal{T})$ locations, as well as the maximal Lyapunov exponents (mLEs), were computed. The dependencies of $\mathcal{A}$ and $\mathcal{T}$ on $r$, the bifurcation diagram, the $\mathcal{A}-\mathcal{T}$ plane, and the relation between $\mathcal{A}$ and mLE, are shown in Fig.~\ref{logistic}. When chaos is most developed ($r=4$), the trajectories approach the point $(\mathcal{A},\mathcal{T}) = (1,2/3)$, identical for white noise [Fig.~\ref{logistic}(d)]. Hints that fully developed chaos behaves this way were noted in case of the Lorenz system \citep{tarnopolski16,zunino17}. With increasing $r$, a gradual decrease in $\mathcal{T}$ occurs, with wells in periodic windows [Fig.~\ref{logistic}(b)]. Note that $\mathcal{T}=1$ up to $r\lesssim 3.7$, because apparently the orbits, even when chaotic, are strictly alternating. $\mathcal{A}$ is more sensitive to changes in dynamics, as $\mathcal{A}=2$ for period-2 orbits  (see the beginning of Sect.~\ref{sect2.2}) before the bifurcation at $r=1+\sqrt{6}\approx 3.45$, and then systematically decreases in the period-4 window before the next bifurcation at $r\approx 3.54$ [Fig.~\ref{logistic}(a)]. There are also shallow wells at periodic windows within the chaotic zone. The path in the $\mathcal{A}-\mathcal{T}$ plane is jagged, and various changes in dynamics are manifested, corresponding to changes in mLEs [Fig.~\ref{logistic}(e)] and in the bifurcation diagram [Fig.~\ref{logistic}(c)].

A tight, positive correlation between mLE and $H$ for the Chirikov map was discovered \citep{tarnopolski18} (see also \citep{steeb05}). Differences between (quasi)periodic and chaotic systems were observed in the coarse-grained sequences in the $\mathcal{A}-\mathcal{T}$ plane in case of a sinusoidally driven thermostat \citep{zhao18}. These connections between chaos, $H$, and the $\mathcal{A}-\mathcal{T}$ plane are nontrivial, and require further research. Describing a universal (if existent) behavior of chaotic systems in the $\mathcal{A}-\mathcal{T}$ plane in an interesting perspective. The possibility of differentiating between chaotic and stochastic time series is a hopefully attainable application. 

\subsection{Markets}
\label{markets}

The efficient market hypothesis \citep{Mantegna2000} is a key concept in finance. If a market exhibits $H\neq 0.5$, then it might allow arbitrage. A classification of developed, emerging and frontier stock markets in the $\mathcal{A}-\mathcal{T}$ plane was performed in \citep{zunino17}, and showed that these three categories of markets occupy distinct regions of the $(\mathcal{A},\mathcal{T})$ space. Hence, the $\mathcal{A}-\mathcal{T}$ plane is a useful tool for classifying data with different underlying dynamics. 

\begin{figure}
\centering
\includegraphics[width=0.7\columnwidth]{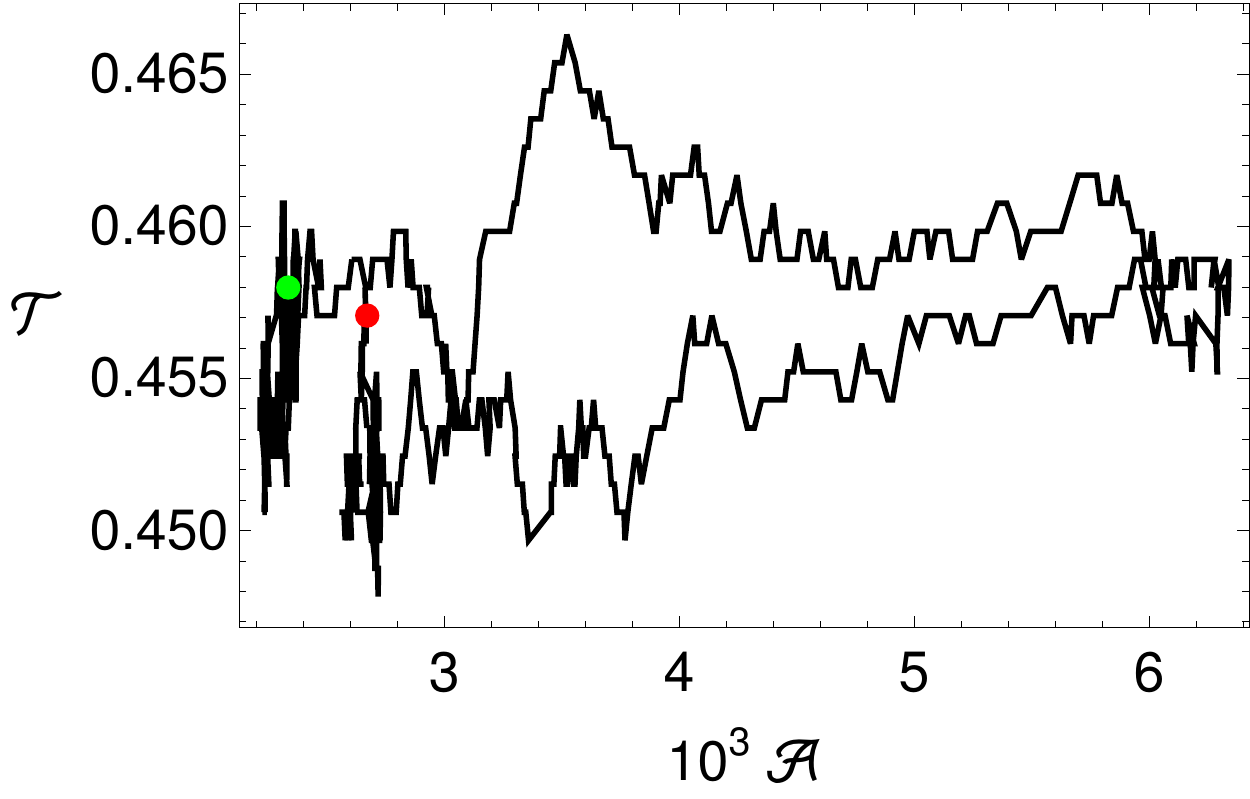}
\caption{Time evolution of the WIG. The green dot marks the start point, and the red dot denotes the end point. Note the scale of the horizontal axis. See text for details. }
\label{wig}
\end{figure}

The time evolution of $H$ of many stock markets shows it is oscillating around $H=0.5$ \citep{carbone04}. Similarly, one can investigate how $\mathcal{A}$ and $\mathcal{T}$ evolve, and hence trace their values in the $\mathcal{A}-\mathcal{T}$ plane at different times. As an example consider the Warsaw Stock Exchange Index (Warszawski Indeks Gie\l dowy, WIG) \footnote{\url{https://www.investing.com/indices/wig-historical-data}, accessed on November 10, 2019}. Its location in the $\mathcal{A}-\mathcal{T}$ plane, right on the fBm line, is depicted in Fig.~\ref{applications}, and the estimates of $H$, obtained by solving Eq.~(\ref{eq1}) and (\ref{eq30}) with $n=2166$, yield $H\approx 0.49-0.59$. The DWT method returns $H=0.48$. The time evolution is depicted in Fig.~\ref{wig}. It is obtained by partitioning the whole time series into overlapping segments of size $n/2$, advancing each segment by one point. Throughout its history, the WIG remains confined in a small region of the $\mathcal{A}-\mathcal{T}$ plane.


\subsection{Active galactic nuclei}
\label{agn}

A core focus in astronomy is the investigation of apparent variability of various celestial objects such as asteroids, stars, or galaxies. Within the latter, of particular interest are active galactic nuclei (AGNs), further divided into several types \citep{Urry95}. Among them, blazars are peculiar AGNs pointing their relativistic jets toward the Earth. Blazars are commonly divided further into two subgroups, i.e. flat spectrum radio quasars (FSRQs) and BL Lacertae (BL Lac) objects, based on characteristics visible in their optical spectra. In a recent study \citep{zywucka19} it was found that faint FSRQ and BL Lac candidates located behind the Magellanic Clouds are clearly separated in the $\mathcal{A}-\mathcal{T}$ plane, with means of $\mathcal{A}\approx 0.3$ and $\mathcal{A}\approx 0.7$, respectively. This differentiation, based solely on the temporal data in the form of light curves, is another proof that employing the $\mathcal{A}-\mathcal{T}$ plane as a classification tool is a promising approach.

\subsection{Other}

The worked-out examples from Sect.~\ref{bacteria}--\ref{agn} do not exhaust the possible applications of the $\mathcal{A}-\mathcal{T}$ plane in regard to constraining the value of $H$, or classifying time series. Some other interesting instances include, but are not restricted to: ferro- and paramagnetic states of the Heisenberg model that exhibit $H\sim 1$ and $H\sim 0.5$, respectively \citep{zhang93}, and should be easily distinguishable in the $\mathcal{A}-\mathcal{T}$ plane; a photonic integrated circuit yields $0.2\lesssim H\lesssim 0.8$ for varied electric field of the feedback, coupled with chaotic behavior \citep{chlouverakis08}; cataclysmic variable stars observed in X-rays exhibit long-term memory, $H>0.5$, suggesting the accretion is driven by magnetic fields \citep{anzolin10}; football matches can follow the rules of fBm with $H\sim 0.7$ \citep{kijima14}; persistence of amoeboid motion \citep{makarava14} as well as {\it Nitzschia} sp. diatoms \citep{marguia15}; Solar wind proton density fluctuations are characterized by $H\sim 0.8$, placing constraints on the models of kinetic turbulence \citep{carbone18}; values $H>0.5$ were computed for epileptic patients' brain activity, quantified via magnetoencephalographic recordings, and appear to be a promising additional diagnostic tool for identifying epileptogenic zones in presurgical evaluation \citep{witton19}. Recall that epilepsy has been already investigated in the $\mathcal{A}-\mathcal{T}$ plane as well \citep{zunino17}.

\section{Summary and open questions}
\label{sect3}

Exact analytical descriptions for the locations of fBm and fGn in the $\mathcal{A}-\mathcal{T}$ plane were derived. Working approximations were also obtained in the following forms:
\begin{eqnarray}
\left\{ \begin{array}{ll}
\mathcal{A}_{\rm fBm}(H,n) &= (H+1)(2H+1)n^{-2H} \\
\mathcal{T}_{\rm fBm}(H) &= 1-\frac{2}{\pi}\arcsin \left( 2^{H-1} \right) \\
\end{array} \right.
\label{}
\end{eqnarray}
and
\begin{eqnarray}
\left\{ \begin{array}{l}
\mathcal{A}_{\rm fGn}(H) = 2 - 2^{2H-1} \\
\mathcal{T}_{\rm fGn}(H) = 1-\frac{2}{\pi}\arcsin\left( \frac{1}{2}\sqrt{\frac{3^{2H}-2^{2H+1}-1}{2^{2H}-4}} \right) \\
\end{array} \right.
\label{}
\end{eqnarray}
and were demonstrated to be adequate for time series with any reasonable length $n$. This allows to classify time series of any length, respective to fBm and fGn, without relying on time-consuming numerical simulations. These analyses add to the theoretical results regarding fBm and fGn \citep{zunino08,Sadhu18}. The same methodology was applied to ARMA$(p,q)$ processes. Analytic descriptions of the available regions of the $\mathcal{A}-\mathcal{T}$ plane were derived and illustrated for $p+q\leq 2$.

Further research on $\mathcal{A}$ is required, as it has been rarely utilized, with some recent, nonextensive examples in astronomy \citep{Shin09,mowlavi,sokolovsky16,perez17,zywucka19} (but see also \citep{lafler65}). The interrelations between ordinal patterns, persistence, and chaos \citep{zanin12,ribeiro17} are linked even tighter with the bidimensional scheme of the $\mathcal{A}-\mathcal{T}$ plane. The presented methodology is valid for any Gaussian process, but it should be emphasized that the locations $(\mathcal{A},\mathcal{T})$ can be computed for arbitrary time series, serving e.g. as classification or clustering methods for empirical data.

Naturally, a question about $\mathcal{A}-\mathcal{T}$ representations of other stochastic processes arises. Examples include the following:
\begin{itemize}
\item Representation of colored noise, i.e. power spectral densities (PSDs) of the form $1/f^\beta$. fBm can be associated with $\beta\in (1,3)$, and fGn is characterized by $\beta\in (-1,1)$. However, power laws are ubiquituous in nature, hence their locations in the $\mathcal{A}-\mathcal{T}$ plane, for any $\beta$, is an interesting and challenging problem.
\item In some fields, e.g. in astronomy, the observed signals often yield PSDs of the form $1/f^\beta+C$, where $C$ is the so called Poisson noise level, coming from the statistical noise due to uncertainties in the measurements; above a certain frequency, the PSD transitions from a power law to white noise. A representation of such processes in the $\mathcal{A}-\mathcal{T}$ plane is crucial in classifying light curves of several sources, e.g. AGN.
\item Other continuous-time models, e.g. continuous ARMA \citep{Brockwell2014}, have been developed. The simplest in this family is the OU process, which is a continuous analog of the AR(1) process, with the same $\mathcal{A}-\mathcal{T}$ representation. Introducing long-term memory leads to continuous autoregressive fractionally integrated moving average models \citep{Tsai09}.
\end{itemize}

\begin{acknowledgments}
The author thanks Ido Golding for providing the data of the mRNA motion inside the bacteria. Support by the Polish National Science Center through OPUS Grant No. 2017/25/B/ST9/01208 is acknowledged.
\end{acknowledgments}

\appendix*

\section{Ornstein-Uhlenbeck process}
\label{app}

The OU process with mean $\mu$ is given by the stochastic differential equation
\begin{equation}
\d x_t = \theta\left(\mu - x_t\right) \d t + \sigma\d\varepsilon_t,
\label{A1}
\end{equation}
where $\theta>0$ and $\sigma>0$. The correlation function for lag $d$ is
\begin{equation}
\rho_d = \exp\left(-\theta d\right),
\label{A2}
\end{equation}
which via Eq.~(\ref{eq40}) and (\ref{eq38}) leads to
\begin{equation}
\mathcal{T}_{\rm OU}\left(\theta\right) = 1-\frac{2}{\pi}\arcsin\left( \frac{1}{2}\sqrt{1+\exp\left( -\theta \right)} \right),
\label{A3}
\end{equation}
reaching its minimum of $1/2$ for $\theta=0$, and its maximum of $2/3$ when $\theta\to\infty$.

The covariance function
\begin{equation}
E\left[ x_t x_s \right] = \frac{\sigma^2}{2\theta} \exp\left( -\theta|t-s| \right)
\label{A4}
\end{equation}
gives via Eq.~(\ref{eq38true}) and (\ref{eq41}) and on simplification
\begin{equation}
\mathcal{A}_{\rm OU}\left(\theta\right) = 1 + \sinh\theta - \cosh\theta,
\label{A5}
\end{equation}
reaching its minimum of $0$ for $\theta=0$, and its maximum of $1$ when $\theta\to\infty$. Eq.~(\ref{A5}) can be solved for $\theta$ and inserted into Eq.~(\ref{A3}), which yields Eq.~(\ref{eq47}), valid for $\mathcal{A}\in[0,1]$.

\bibliography{mybibfile}

\begin{thebibliography}{68}%
\makeatletter
\providecommand \@ifxundefined [1]{%
 \@ifx{#1\undefined}
}%
\providecommand \@ifnum [1]{%
 \ifnum #1\expandafter \@firstoftwo
 \else \expandafter \@secondoftwo
 \fi
}%
\providecommand \@ifx [1]{%
 \ifx #1\expandafter \@firstoftwo
 \else \expandafter \@secondoftwo
 \fi
}%
\providecommand \natexlab [1]{#1}%
\providecommand \enquote  [1]{``#1''}%
\providecommand \bibnamefont  [1]{#1}%
\providecommand \bibfnamefont [1]{#1}%
\providecommand \citenamefont [1]{#1}%
\providecommand \href@noop [0]{\@secondoftwo}%
\providecommand \href [0]{\begingroup \@sanitize@url \@href}%
\providecommand \@href[1]{\@@startlink{#1}\@@href}%
\providecommand \@@href[1]{\endgroup#1\@@endlink}%
\providecommand \@sanitize@url [0]{\catcode `\\12\catcode `\$12\catcode
  `\&12\catcode `\#12\catcode `\^12\catcode `\_12\catcode `\%12\relax}%
\providecommand \@@startlink[1]{}%
\providecommand \@@endlink[0]{}%
\providecommand \url  [0]{\begingroup\@sanitize@url \@url }%
\providecommand \@url [1]{\endgroup\@href {#1}{\urlprefix }}%
\providecommand \urlprefix  [0]{URL }%
\providecommand \Eprint [0]{\href }%
\providecommand \doibase [0]{http://dx.doi.org/}%
\providecommand \selectlanguage [0]{\@gobble}%
\providecommand \bibinfo  [0]{\@secondoftwo}%
\providecommand \bibfield  [0]{\@secondoftwo}%
\providecommand \translation [1]{[#1]}%
\providecommand \BibitemOpen [0]{}%
\providecommand \bibitemStop [0]{}%
\providecommand \bibitemNoStop [0]{.\EOS\space}%
\providecommand \EOS [0]{\spacefactor3000\relax}%
\providecommand \BibitemShut  [1]{\csname bibitem#1\endcsname}%
\let\auto@bib@innerbib\@empty
\bibitem [{\citenamefont {{Beran}}(1994)}]{beran}%
  \BibitemOpen
  \bibfield  {author} {\bibinfo {author} {\bibfnamefont {J.}~\bibnamefont
  {{Beran}}},\ }\href@noop {} {\emph {\bibinfo {title} {{Statistics for Long
  Memory Processes}}}}\ (\bibinfo  {publisher} {Chapman and Hall, New York},\
  \bibinfo {year} {1994})\BibitemShut {NoStop}%
\bibitem [{\citenamefont {{Brockwell}}\ and\ \citenamefont
  {{Davis}}(1996)}]{brockwell}%
  \BibitemOpen
  \bibfield  {author} {\bibinfo {author} {\bibfnamefont {P.~J.}\ \bibnamefont
  {{Brockwell}}}\ and\ \bibinfo {author} {\bibfnamefont {R.~A.}\ \bibnamefont
  {{Davis}}},\ }\href {\doibase 10.1007/978-1-4419-0320-4} {\emph {\bibinfo
  {title} {{Time Series: Theory and Methods, 2nd ed.}}}}\ (\bibinfo
  {publisher} {Springer-Verlag New York},\ \bibinfo {year} {1996})\BibitemShut
  {NoStop}%
\bibitem [{\citenamefont {{Brockwell}}\ and\ \citenamefont
  {{Davis}}(2002)}]{brockwell2}%
  \BibitemOpen
  \bibfield  {author} {\bibinfo {author} {\bibfnamefont {P.~J.}\ \bibnamefont
  {{Brockwell}}}\ and\ \bibinfo {author} {\bibfnamefont {R.~A.}\ \bibnamefont
  {{Davis}}},\ }\href {\doibase 10.1007/b97391} {\emph {\bibinfo {title}
  {{Introduction to Time Series and Forecasting, 2nd ed.}}}}\ (\bibinfo
  {publisher} {Springer-Verlag New York},\ \bibinfo {year} {2002})\BibitemShut
  {NoStop}%
\bibitem [{\citenamefont {{Beran}}\ \emph {et~al.}(2013)\citenamefont
  {{Beran}}, \citenamefont {{Feng}}, \citenamefont {{Ghosh}},\ and\
  \citenamefont {{Kulik}}}]{beran2}%
  \BibitemOpen
  \bibfield  {author} {\bibinfo {author} {\bibfnamefont {J.}~\bibnamefont
  {{Beran}}}, \bibinfo {author} {\bibfnamefont {Y.}~\bibnamefont {{Feng}}},
  \bibinfo {author} {\bibfnamefont {S.}~\bibnamefont {{Ghosh}}}, \ and\
  \bibinfo {author} {\bibfnamefont {R.}~\bibnamefont {{Kulik}}},\ }\href
  {\doibase 10.1007/978-3-642-35512-7} {\emph {\bibinfo {title} {{Long Memory
  Processes: Probabilistic Properties and Statistical Methods}}}}\ (\bibinfo
  {publisher} {Springer-Verlag, Berlin Heidelberg},\ \bibinfo {year}
  {2013})\BibitemShut {NoStop}%
\bibitem [{\citenamefont {{Hegger}}\ \emph {et~al.}(1999)\citenamefont
  {{Hegger}}, \citenamefont {{Kantz}},\ and\ \citenamefont
  {{Schreiber}}}]{hegger99}%
  \BibitemOpen
  \bibfield  {author} {\bibinfo {author} {\bibfnamefont {R.}~\bibnamefont
  {{Hegger}}}, \bibinfo {author} {\bibfnamefont {H.}~\bibnamefont {{Kantz}}}, \
  and\ \bibinfo {author} {\bibfnamefont {T.}~\bibnamefont {{Schreiber}}},\
  }\href {\doibase 10.1063/1.166424} {\bibfield  {journal} {\bibinfo  {journal}
  {Chaos}\ }\textbf {\bibinfo {volume} {9}},\ \bibinfo {pages} {413} (\bibinfo
  {year} {1999})}\BibitemShut {NoStop}%
\bibitem [{\citenamefont {{Bandt}}\ and\ \citenamefont
  {{Pompe}}(2002)}]{bandt_pompe_02}%
  \BibitemOpen
  \bibfield  {author} {\bibinfo {author} {\bibfnamefont {C.}~\bibnamefont
  {{Bandt}}}\ and\ \bibinfo {author} {\bibfnamefont {B.}~\bibnamefont
  {{Pompe}}},\ }\href {\doibase 10.1103/PhysRevLett.88.174102} {\bibfield
  {journal} {\bibinfo  {journal} {Physical Review Letters}\ }\textbf {\bibinfo
  {volume} {88}},\ \bibinfo {pages} {174102} (\bibinfo {year}
  {2002})}\BibitemShut {NoStop}%
\bibitem [{\citenamefont {{Maggs}}\ and\ \citenamefont
  {{Morales}}(2013)}]{maggs13}%
  \BibitemOpen
  \bibfield  {author} {\bibinfo {author} {\bibfnamefont {J.~E.}\ \bibnamefont
  {{Maggs}}}\ and\ \bibinfo {author} {\bibfnamefont {G.~J.}\ \bibnamefont
  {{Morales}}},\ }\href {\doibase 10.1088/0741-3335/55/8/085015} {\bibfield
  {journal} {\bibinfo  {journal} {Plasma Physics and Controlled Fusion}\
  }\textbf {\bibinfo {volume} {55}},\ \bibinfo {eid} {085015} (\bibinfo {year}
  {2013})}\BibitemShut {NoStop}%
\bibitem [{\citenamefont {{Ribeiro}}\ \emph {et~al.}(2017)\citenamefont
  {{Ribeiro}}, \citenamefont {{Jauregui}}, \citenamefont {{Zunino}},\ and\
  \citenamefont {{Lenzi}}}]{ribeiro17}%
  \BibitemOpen
  \bibfield  {author} {\bibinfo {author} {\bibfnamefont {H.~V.}\ \bibnamefont
  {{Ribeiro}}}, \bibinfo {author} {\bibfnamefont {M.}~\bibnamefont
  {{Jauregui}}}, \bibinfo {author} {\bibfnamefont {L.}~\bibnamefont
  {{Zunino}}}, \ and\ \bibinfo {author} {\bibfnamefont {E.~K.}\ \bibnamefont
  {{Lenzi}}},\ }\href {\doibase 10.1103/PhysRevE.95.062106} {\bibfield
  {journal} {\bibinfo  {journal} {Physical Review E}\ }\textbf {\bibinfo
  {volume} {95}},\ \bibinfo {eid} {062106} (\bibinfo {year}
  {2017})}\BibitemShut {NoStop}%
\bibitem [{\citenamefont {{Simonsen}}\ \emph {et~al.}(1998)\citenamefont
  {{Simonsen}}, \citenamefont {{Hansen}},\ and\ \citenamefont
  {{Nes}}}]{simonsen}%
  \BibitemOpen
  \bibfield  {author} {\bibinfo {author} {\bibfnamefont {I.}~\bibnamefont
  {{Simonsen}}}, \bibinfo {author} {\bibfnamefont {A.}~\bibnamefont
  {{Hansen}}}, \ and\ \bibinfo {author} {\bibfnamefont {O.~M.}\ \bibnamefont
  {{Nes}}},\ }\href {\doibase 10.1103/PhysRevE.58.2779} {\bibfield  {journal}
  {\bibinfo  {journal} {Physical Review E}\ }\textbf {\bibinfo {volume} {58}},\
  \bibinfo {pages} {2779} (\bibinfo {year} {1998})},\ \Eprint
  {http://arxiv.org/abs/cond-mat/9707153} {cond-mat/9707153} \BibitemShut
  {NoStop}%
\bibitem [{\citenamefont {{Carbone}}\ \emph {et~al.}(2004)\citenamefont
  {{Carbone}}, \citenamefont {{Castelli}},\ and\ \citenamefont
  {{Stanley}}}]{carbone04}%
  \BibitemOpen
  \bibfield  {author} {\bibinfo {author} {\bibfnamefont {A.}~\bibnamefont
  {{Carbone}}}, \bibinfo {author} {\bibfnamefont {G.}~\bibnamefont
  {{Castelli}}}, \ and\ \bibinfo {author} {\bibfnamefont {H.~E.}\ \bibnamefont
  {{Stanley}}},\ }\href {\doibase 10.1016/j.physa.2004.06.130} {\bibfield
  {journal} {\bibinfo  {journal} {Physica A: Statistical Mechanics and its
  Applications}\ }\textbf {\bibinfo {volume} {344}},\ \bibinfo {pages} {267}
  (\bibinfo {year} {2004})}\BibitemShut {NoStop}%
\bibitem [{\citenamefont {{Arianos}}\ and\ \citenamefont
  {{Carbone}}(2007)}]{arianos07}%
  \BibitemOpen
  \bibfield  {author} {\bibinfo {author} {\bibfnamefont {S.}~\bibnamefont
  {{Arianos}}}\ and\ \bibinfo {author} {\bibfnamefont {A.}~\bibnamefont
  {{Carbone}}},\ }\href {\doibase 10.1016/j.physa.2007.02.074} {\bibfield
  {journal} {\bibinfo  {journal} {Physica A: Statistical Mechanics and its
  Applications}\ }\textbf {\bibinfo {volume} {382}},\ \bibinfo {pages} {9}
  (\bibinfo {year} {2007})}\BibitemShut {NoStop}%
\bibitem [{\citenamefont {{Lacasa}}\ and\ \citenamefont
  {{Toral}}(2010)}]{lacasa10}%
  \BibitemOpen
  \bibfield  {author} {\bibinfo {author} {\bibfnamefont {L.}~\bibnamefont
  {{Lacasa}}}\ and\ \bibinfo {author} {\bibfnamefont {R.}~\bibnamefont
  {{Toral}}},\ }\href {\doibase 10.1103/PhysRevE.82.036120} {\bibfield
  {journal} {\bibinfo  {journal} {Physical Review E}\ }\textbf {\bibinfo
  {volume} {82}},\ \bibinfo {eid} {036120} (\bibinfo {year}
  {2010})}\BibitemShut {NoStop}%
\bibitem [{\citenamefont {{Tarnopolski}}(2018)}]{tarnopolski18}%
  \BibitemOpen
  \bibfield  {author} {\bibinfo {author} {\bibfnamefont {M.}~\bibnamefont
  {{Tarnopolski}}},\ }\href {\doibase 10.1016/j.physa.2017.08.159} {\bibfield
  {journal} {\bibinfo  {journal} {Physica A: Statistical Mechanics and its
  Applications}\ }\textbf {\bibinfo {volume} {490}},\ \bibinfo {pages} {834}
  (\bibinfo {year} {2018})}\BibitemShut {NoStop}%
\bibitem [{\citenamefont {{Tarnopolski}}(2016)}]{tarnopolski16}%
  \BibitemOpen
  \bibfield  {author} {\bibinfo {author} {\bibfnamefont {M.}~\bibnamefont
  {{Tarnopolski}}},\ }\href {\doibase 10.1016/j.physa.2016.06.004} {\bibfield
  {journal} {\bibinfo  {journal} {Physica A: Statistical Mechanics and its
  Applications}\ }\textbf {\bibinfo {volume} {461}},\ \bibinfo {pages} {662}
  (\bibinfo {year} {2016})}\BibitemShut {NoStop}%
\bibitem [{\citenamefont {{Peng}}\ \emph {et~al.}(1994)\citenamefont {{Peng}},
  \citenamefont {{Buldyrev}}, \citenamefont {{Havlin}}, \citenamefont
  {{Simons}}, \citenamefont {{Stanley}},\ and\ \citenamefont
  {{Goldberger}}}]{peng94}%
  \BibitemOpen
  \bibfield  {author} {\bibinfo {author} {\bibfnamefont {C.-K.}\ \bibnamefont
  {{Peng}}}, \bibinfo {author} {\bibfnamefont {S.~V.}\ \bibnamefont
  {{Buldyrev}}}, \bibinfo {author} {\bibfnamefont {S.}~\bibnamefont
  {{Havlin}}}, \bibinfo {author} {\bibfnamefont {M.}~\bibnamefont {{Simons}}},
  \bibinfo {author} {\bibfnamefont {H.~E.}\ \bibnamefont {{Stanley}}}, \ and\
  \bibinfo {author} {\bibfnamefont {A.~L.}\ \bibnamefont {{Goldberger}}},\
  }\href {\doibase 10.1103/PhysRevE.49.1685} {\bibfield  {journal} {\bibinfo
  {journal} {Physical Review E}\ }\textbf {\bibinfo {volume} {49}},\ \bibinfo
  {pages} {1685} (\bibinfo {year} {1994})}\BibitemShut {NoStop}%
\bibitem [{\citenamefont {{Peng}}\ \emph {et~al.}(1995)\citenamefont {{Peng}},
  \citenamefont {{Havlin}}, \citenamefont {{Stanley}},\ and\ \citenamefont
  {{Goldberger}}}]{peng95}%
  \BibitemOpen
  \bibfield  {author} {\bibinfo {author} {\bibfnamefont {C.-K.}\ \bibnamefont
  {{Peng}}}, \bibinfo {author} {\bibfnamefont {S.}~\bibnamefont {{Havlin}}},
  \bibinfo {author} {\bibfnamefont {H.~E.}\ \bibnamefont {{Stanley}}}, \ and\
  \bibinfo {author} {\bibfnamefont {A.~L.}\ \bibnamefont {{Goldberger}}},\
  }\href {\doibase 10.1063/1.166141} {\bibfield  {journal} {\bibinfo  {journal}
  {Chaos}\ }\textbf {\bibinfo {volume} {5}},\ \bibinfo {pages} {82} (\bibinfo
  {year} {1995})}\BibitemShut {NoStop}%
\bibitem [{\citenamefont {{Zunino}}\ \emph
  {et~al.}(2017{\natexlab{a}})\citenamefont {{Zunino}}, \citenamefont
  {{Olivares}}, \citenamefont {{Bariviera}},\ and\ \citenamefont
  {{Rosso}}}]{zunino17}%
  \BibitemOpen
  \bibfield  {author} {\bibinfo {author} {\bibfnamefont {L.}~\bibnamefont
  {{Zunino}}}, \bibinfo {author} {\bibfnamefont {F.}~\bibnamefont
  {{Olivares}}}, \bibinfo {author} {\bibfnamefont {A.~F.}\ \bibnamefont
  {{Bariviera}}}, \ and\ \bibinfo {author} {\bibfnamefont {O.~A.}\ \bibnamefont
  {{Rosso}}},\ }\href {\doibase 10.1016/j.physleta.2017.01.047} {\bibfield
  {journal} {\bibinfo  {journal} {Physics Letters A}\ }\textbf {\bibinfo
  {volume} {381}},\ \bibinfo {pages} {1021} (\bibinfo {year}
  {2017}{\natexlab{a}})}\BibitemShut {NoStop}%
\bibitem [{\citenamefont {{Zhao}}\ and\ \citenamefont
  {{Morales}}(2018)}]{zhao18}%
  \BibitemOpen
  \bibfield  {author} {\bibinfo {author} {\bibfnamefont {Y.}~\bibnamefont
  {{Zhao}}}\ and\ \bibinfo {author} {\bibfnamefont {G.~J.}\ \bibnamefont
  {{Morales}}},\ }\href {\doibase 10.1103/PhysRevE.98.022213} {\bibfield
  {journal} {\bibinfo  {journal} {Physical Review E}\ }\textbf {\bibinfo
  {volume} {98}},\ \bibinfo {eid} {022213} (\bibinfo {year}
  {2018})}\BibitemShut {NoStop}%
\bibitem [{\citenamefont {{Zunino}}\ \emph
  {et~al.}(2017{\natexlab{b}})\citenamefont {{Zunino}}, \citenamefont
  {{Olivares}}, \citenamefont {{Scholkmann}},\ and\ \citenamefont
  {{Rosso}}}]{zunino2017}%
  \BibitemOpen
  \bibfield  {author} {\bibinfo {author} {\bibfnamefont {L.}~\bibnamefont
  {{Zunino}}}, \bibinfo {author} {\bibfnamefont {F.}~\bibnamefont
  {{Olivares}}}, \bibinfo {author} {\bibfnamefont {F.}~\bibnamefont
  {{Scholkmann}}}, \ and\ \bibinfo {author} {\bibfnamefont {O.~A.}\
  \bibnamefont {{Rosso}}},\ }\href {\doibase 10.1016/j.physleta.2017.03.052}
  {\bibfield  {journal} {\bibinfo  {journal} {Physics Letters A}\ }\textbf
  {\bibinfo {volume} {381}},\ \bibinfo {pages} {1883} (\bibinfo {year}
  {2017}{\natexlab{b}})}\BibitemShut {NoStop}%
\bibitem [{\citenamefont {{Traversaro}}\ \emph {et~al.}(2018)\citenamefont
  {{Traversaro}}, \citenamefont {{Redelico}}, \citenamefont {{Risk}},
  \citenamefont {{Frery}},\ and\ \citenamefont {{Rosso}}}]{traversaro18}%
  \BibitemOpen
  \bibfield  {author} {\bibinfo {author} {\bibfnamefont {F.}~\bibnamefont
  {{Traversaro}}}, \bibinfo {author} {\bibfnamefont {F.~O.}\ \bibnamefont
  {{Redelico}}}, \bibinfo {author} {\bibfnamefont {M.~R.}\ \bibnamefont
  {{Risk}}}, \bibinfo {author} {\bibfnamefont {A.~C.}\ \bibnamefont {{Frery}}},
  \ and\ \bibinfo {author} {\bibfnamefont {O.~A.}\ \bibnamefont {{Rosso}}},\
  }\href {\doibase 10.1063/1.5022021} {\bibfield  {journal} {\bibinfo
  {journal} {Chaos}\ }\textbf {\bibinfo {volume} {28}},\ \bibinfo {eid}
  {075502} (\bibinfo {year} {2018})}\BibitemShut {NoStop}%
\bibitem [{\citenamefont {{Bandt}}\ and\ \citenamefont
  {{Shiha}}(2007)}]{bandt07}%
  \BibitemOpen
  \bibfield  {author} {\bibinfo {author} {\bibfnamefont {C.}~\bibnamefont
  {{Bandt}}}\ and\ \bibinfo {author} {\bibfnamefont {F.}~\bibnamefont
  {{Shiha}}},\ }\href {\doibase 10.1111/j.1467-9892.2007.00528.x} {\bibfield
  {journal} {\bibinfo  {journal} {Journal of Time Series Analysis}\ }\textbf
  {\bibinfo {volume} {28}},\ \bibinfo {pages} {646} (\bibinfo {year}
  {2007})}\BibitemShut {NoStop}%
\bibitem [{Note1()}]{Note1}%
  \BibitemOpen
  \bibinfo {note} {In the derivation of $\protect \mathcal {T}_{\protect \rm
  fGn}$ and $\protect \mathcal {T}_{\protect \rm DfGn}$ one can actually employ
  only the relations for fBm; the calculations would then become quite
  burdensome, though.}\BibitemShut {Stop}%
\bibitem [{\citenamefont {{Kendall}}\ and\ \citenamefont
  {{Stuart}}(1973)}]{kendall}%
  \BibitemOpen
  \bibfield  {author} {\bibinfo {author} {\bibfnamefont {M.}~\bibnamefont
  {{Kendall}}}\ and\ \bibinfo {author} {\bibfnamefont {A.}~\bibnamefont
  {{Stuart}}},\ }\href@noop {} {\emph {\bibinfo {title} {{The advanced theory
  of statistics}}}}\ (\bibinfo  {publisher} {London: Griffin, 3rd ed.},\
  \bibinfo {year} {1973})\BibitemShut {NoStop}%
\bibitem [{Note2()}]{Note2}%
  \BibitemOpen
  \bibinfo {note} {The first and last points cannot form a turning point, hence
  the subtraction of 2 in $\mu _T$.}\BibitemShut {Stop}%
\bibitem [{\citenamefont {{Sinn}}\ and\ \citenamefont
  {{Keller}}(2008)}]{sinn08}%
  \BibitemOpen
  \bibfield  {author} {\bibinfo {author} {\bibfnamefont {M.}~\bibnamefont
  {{Sinn}}}\ and\ \bibinfo {author} {\bibfnamefont {K.}~\bibnamefont
  {{Keller}}},\ }\href@noop {} {\bibfield  {journal} {\bibinfo  {journal}
  {arXiv e-prints}\ ,\ \bibinfo {eid} {arXiv:0801.1598}} (\bibinfo {year}
  {2008})},\ \Eprint {http://arxiv.org/abs/0801.1598} {arXiv:0801.1598
  [math.PR]} \BibitemShut {NoStop}%
\bibitem [{\citenamefont {{von Neumann}}(1941{\natexlab{a}})}]{neumann}%
  \BibitemOpen
  \bibfield  {author} {\bibinfo {author} {\bibfnamefont {J.}~\bibnamefont {{von
  Neumann}}},\ }\href {\doibase 10.1214/aoms/1177731746} {\bibfield  {journal}
  {\bibinfo  {journal} {The Annals of Mathematical Statistics}\ }\textbf
  {\bibinfo {volume} {12}},\ \bibinfo {pages} {153} (\bibinfo {year}
  {1941}{\natexlab{a}})}\BibitemShut {NoStop}%
\bibitem [{\citenamefont {{von Neumann}}(1941{\natexlab{b}})}]{neumann2}%
  \BibitemOpen
  \bibfield  {author} {\bibinfo {author} {\bibfnamefont {J.}~\bibnamefont {{von
  Neumann}}},\ }\href {\doibase 10.1214/aoms/1177731677} {\bibfield  {journal}
  {\bibinfo  {journal} {The Annals of Mathematical Statistics}\ }\textbf
  {\bibinfo {volume} {12}},\ \bibinfo {pages} {367} (\bibinfo {year}
  {1941}{\natexlab{b}})}\BibitemShut {NoStop}%
\bibitem [{\citenamefont {{Kendall}}(1971)}]{kendall1971}%
  \BibitemOpen
  \bibfield  {author} {\bibinfo {author} {\bibfnamefont {M.~G.}\ \bibnamefont
  {{Kendall}}},\ }\href {\doibase 10.2307/2334525} {\bibfield  {journal}
  {\bibinfo  {journal} {Biometrika}\ }\textbf {\bibinfo {volume} {58}},\
  \bibinfo {pages} {369} (\bibinfo {year} {1971})}\BibitemShut {NoStop}%
\bibitem [{\citenamefont {{Mowlavi}}(2014)}]{mowlavi}%
  \BibitemOpen
  \bibfield  {author} {\bibinfo {author} {\bibfnamefont {N.}~\bibnamefont
  {{Mowlavi}}},\ }\href {\doibase 10.1051/0004-6361/201322648} {\bibfield
  {journal} {\bibinfo  {journal} {Astronomy \& Astrophysics}\ }\textbf
  {\bibinfo {volume} {568}},\ \bibinfo {eid} {A78} (\bibinfo {year}
  {2014})}\BibitemShut {NoStop}%
\bibitem [{\citenamefont {{Williams}}(1941)}]{williams}%
  \BibitemOpen
  \bibfield  {author} {\bibinfo {author} {\bibfnamefont {J.~D.}\ \bibnamefont
  {{Williams}}},\ }\href {\doibase 10.1214/aoms/1177731756} {\bibfield
  {journal} {\bibinfo  {journal} {The Annals of Mathematical Statistics}\
  }\textbf {\bibinfo {volume} {12}},\ \bibinfo {pages} {239} (\bibinfo {year}
  {1941})}\BibitemShut {NoStop}%
\bibitem [{\citenamefont {{Bingham}}\ and\ \citenamefont
  {{Nelson}}(1981)}]{bingham81}%
  \BibitemOpen
  \bibfield  {author} {\bibinfo {author} {\bibfnamefont {C.}~\bibnamefont
  {{Bingham}}}\ and\ \bibinfo {author} {\bibfnamefont {L.~S.}\ \bibnamefont
  {{Nelson}}},\ }\href {\doibase 10.1080/00401706.1981.10487651} {\bibfield
  {journal} {\bibinfo  {journal} {Technometrics}\ }\textbf {\bibinfo {volume}
  {23}},\ \bibinfo {pages} {285} (\bibinfo {year} {1981})}\BibitemShut
  {NoStop}%
\bibitem [{\citenamefont {{Bartels}}(1982)}]{bartels82}%
  \BibitemOpen
  \bibfield  {author} {\bibinfo {author} {\bibfnamefont {R.}~\bibnamefont
  {{Bartels}}},\ }\href {\doibase 10.1080/01621459.1982.10477764} {\bibfield
  {journal} {\bibinfo  {journal} {Journal of the American Statistical
  Association}\ }\textbf {\bibinfo {volume} {77}},\ \bibinfo {pages} {40 }
  (\bibinfo {year} {1982})}\BibitemShut {NoStop}%
\bibitem [{\citenamefont {Mateus}\ and\ \citenamefont
  {Caeiro}(2018)}]{mateus18}%
  \BibitemOpen
  \bibfield  {author} {\bibinfo {author} {\bibfnamefont {A.}~\bibnamefont
  {Mateus}}\ and\ \bibinfo {author} {\bibfnamefont {F.}~\bibnamefont
  {Caeiro}},\ }\enquote {\bibinfo {title} {Exact and approximate probabilities
  for the null distribution of bartels randomness test},}\ in\ \href {\doibase
  10.1007/978-3-319-76605-8_16} {\emph {\bibinfo {booktitle} {Recent Studies on
  Risk Analysis and Statistical Modeling}}},\ \bibinfo {editor} {edited by\
  \bibinfo {editor} {\bibfnamefont {T.~A.}\ \bibnamefont {Oliveira}}, \bibinfo
  {editor} {\bibfnamefont {C.~P.}\ \bibnamefont {Kitsos}}, \bibinfo {editor}
  {\bibfnamefont {A.}~\bibnamefont {Oliveira}}, \ and\ \bibinfo {editor}
  {\bibfnamefont {L.}~\bibnamefont {Grilo}}}\ (\bibinfo  {publisher} {Springer
  International Publishing},\ \bibinfo {address} {Cham},\ \bibinfo {year}
  {2018})\ pp.\ \bibinfo {pages} {227--240}\BibitemShut {NoStop}%
\bibitem [{\citenamefont {{Deligni\`{e}res}}(2015)}]{delignieres15}%
  \BibitemOpen
  \bibfield  {author} {\bibinfo {author} {\bibfnamefont {D.}~\bibnamefont
  {{Deligni\`{e}res}}},\ }\href {\doibase 10.1155/2015/485623} {\bibfield
  {journal} {\bibinfo  {journal} {Mathematical Problems in Engineering}\
  }\textbf {\bibinfo {volume} {2015}},\ \bibinfo {eid} {485623} (\bibinfo
  {year} {2015})}\BibitemShut {NoStop}%
\bibitem [{Note3()}]{Note3}%
  \BibitemOpen
  \bibinfo {note} {See also \protect \url {www.wolframalpha.com}.}\BibitemShut
  {Stop}%
\bibitem [{\citenamefont {{Apostol}}(1976)}]{apostol}%
  \BibitemOpen
  \bibfield  {author} {\bibinfo {author} {\bibfnamefont {T.~M.}\ \bibnamefont
  {{Apostol}}},\ }\href {\doibase 10.1007/978-1-4757-5579-4} {\emph {\bibinfo
  {title} {{Introduction to Analytic Number Theory}}}}\ (\bibinfo  {publisher}
  {Springer Science+Business Media New York},\ \bibinfo {year}
  {1976})\BibitemShut {NoStop}%
\bibitem [{\citenamefont {{Hasse}}(1930)}]{hasse30}%
  \BibitemOpen
  \bibfield  {author} {\bibinfo {author} {\bibfnamefont {H.}~\bibnamefont
  {{Hasse}}},\ }\href@noop {} {\bibfield  {journal} {\bibinfo  {journal}
  {Mathematische Zeitschrift}\ }\textbf {\bibinfo {volume} {32}},\ \bibinfo
  {pages} {458} (\bibinfo {year} {1930})}\BibitemShut {NoStop}%
\bibitem [{\citenamefont {Golding}\ and\ \citenamefont
  {Cox}(2006)}]{golding06}%
  \BibitemOpen
  \bibfield  {author} {\bibinfo {author} {\bibfnamefont {I.}~\bibnamefont
  {Golding}}\ and\ \bibinfo {author} {\bibfnamefont {E.~C.}\ \bibnamefont
  {Cox}},\ }\href {\doibase 10.1103/PhysRevLett.96.098102} {\bibfield
  {journal} {\bibinfo  {journal} {Physical Review Letters}\ }\textbf {\bibinfo
  {volume} {96}},\ \bibinfo {pages} {098102} (\bibinfo {year}
  {2006})}\BibitemShut {NoStop}%
\bibitem [{\citenamefont {Magdziarz}\ \emph {et~al.}(2009)\citenamefont
  {Magdziarz}, \citenamefont {Weron}, \citenamefont {Burnecki},\ and\
  \citenamefont {Klafter}}]{magdziarz09}%
  \BibitemOpen
  \bibfield  {author} {\bibinfo {author} {\bibfnamefont {M.}~\bibnamefont
  {Magdziarz}}, \bibinfo {author} {\bibfnamefont {A.}~\bibnamefont {Weron}},
  \bibinfo {author} {\bibfnamefont {K.}~\bibnamefont {Burnecki}}, \ and\
  \bibinfo {author} {\bibfnamefont {J.}~\bibnamefont {Klafter}},\ }\href
  {\doibase 10.1103/PhysRevLett.103.180602} {\bibfield  {journal} {\bibinfo
  {journal} {Physical Review Letters}\ }\textbf {\bibinfo {volume} {103}},\
  \bibinfo {pages} {180602} (\bibinfo {year} {2009})}\BibitemShut {NoStop}%
\bibitem [{\citenamefont {Platt}(2015)}]{platt15}%
  \BibitemOpen
  \bibfield  {author} {\bibinfo {author} {\bibfnamefont {D.~J.}\ \bibnamefont
  {Platt}},\ }\href {\doibase https://doi.org/10.1090/S0025-5718-2014-02884-6}
  {\bibfield  {journal} {\bibinfo  {journal} {Mathematics of Computation}\
  }\textbf {\bibinfo {volume} {84}},\ \bibinfo {pages} {1521} (\bibinfo {year}
  {2015})}\BibitemShut {NoStop}%
\bibitem [{Note4()}]{Note4}%
  \BibitemOpen
  \bibinfo {note} {\protect \url
  {https://www.lmfdb.org/zeros/zeta/}}\BibitemShut {NoStop}%
\bibitem [{\citenamefont {Odlyzko}(1987)}]{odlyzko87}%
  \BibitemOpen
  \bibfield  {author} {\bibinfo {author} {\bibfnamefont {A.~M.}\ \bibnamefont
  {Odlyzko}},\ }\href {\doibase
  https://doi.org/10.1090/S0025-5718-1987-0866115-0} {\bibfield  {journal}
  {\bibinfo  {journal} {Mathematics of Computation}\ }\textbf {\bibinfo
  {volume} {48}},\ \bibinfo {pages} {273} (\bibinfo {year} {1987})}\BibitemShut
  {NoStop}%
\bibitem [{\citenamefont {{Montgomery}}(1973)}]{montgomery73}%
  \BibitemOpen
  \bibfield  {author} {\bibinfo {author} {\bibfnamefont {H.~L.}\ \bibnamefont
  {{Montgomery}}},\ }in\ \href@noop {} {\emph {\bibinfo {booktitle} {Analytic
  number theory}}},\ \bibinfo {series} {Proc. Sympos. Pure Math., XXIV,
  Providence, R.I.: American Mathematical Society}, Vol.~\bibinfo {volume}
  {24},\ \bibinfo {editor} {edited by\ \bibinfo {editor} {\bibfnamefont
  {H.~G.}\ \bibnamefont {{Diamond}}}}\ (\bibinfo {year} {1973})\ pp.\ \bibinfo
  {pages} {181--193}\BibitemShut {NoStop}%
\bibitem [{Note5()}]{Note5}%
  \BibitemOpen
  \bibinfo {note} {SILSO World Data Center, Royal Observatory of Belgium,
  Brussels, \protect \url {http://www.sidc.be/silso}}\BibitemShut {NoStop}%
\bibitem [{Note6()}]{Note6}%
  \BibitemOpen
  \bibinfo {note} {Similar value can be obtained with data in \protect \citep
  {tarnopolski16} when the two outliers on the left of Fig.~6b therein are
  discarded.}\BibitemShut {Stop}%
\bibitem [{\citenamefont {Movahed}\ \emph {et~al.}(2006)\citenamefont
  {Movahed}, \citenamefont {Jafari}, \citenamefont {Ghasemi}, \citenamefont
  {Rahvar},\ and\ \citenamefont {Tabar}}]{movahed06}%
  \BibitemOpen
  \bibfield  {author} {\bibinfo {author} {\bibfnamefont {M.~S.}\ \bibnamefont
  {Movahed}}, \bibinfo {author} {\bibfnamefont {G.~R.}\ \bibnamefont {Jafari}},
  \bibinfo {author} {\bibfnamefont {F.}~\bibnamefont {Ghasemi}}, \bibinfo
  {author} {\bibfnamefont {S.}~\bibnamefont {Rahvar}}, \ and\ \bibinfo {author}
  {\bibfnamefont {M.~R.~R.}\ \bibnamefont {Tabar}},\ }\href {\doibase
  10.1088/1742-5468/2006/02/p02003} {\bibfield  {journal} {\bibinfo  {journal}
  {Journal of Statistical Mechanics: Theory and Experiment}\ }\textbf {\bibinfo
  {volume} {2006}},\ \bibinfo {pages} {P02003} (\bibinfo {year}
  {2006})}\BibitemShut {NoStop}%
\bibitem [{\citenamefont {{Steeb}}\ and\ \citenamefont
  {{Andrieu}}(2005)}]{steeb05}%
  \BibitemOpen
  \bibfield  {author} {\bibinfo {author} {\bibfnamefont {W.-H.}\ \bibnamefont
  {{Steeb}}}\ and\ \bibinfo {author} {\bibfnamefont {E.~C.}\ \bibnamefont
  {{Andrieu}}},\ }\href {\doibase 10.1515/zna-2005-0406} {\bibfield  {journal}
  {\bibinfo  {journal} {Zeitschrift Naturforschung Teil A}\ }\textbf {\bibinfo
  {volume} {60}},\ \bibinfo {pages} {252} (\bibinfo {year} {2005})}\BibitemShut
  {NoStop}%
\bibitem [{\citenamefont {Mantegna}\ and\ \citenamefont
  {Stanley}(2000)}]{Mantegna2000}%
  \BibitemOpen
  \bibfield  {author} {\bibinfo {author} {\bibfnamefont {R.~N.}\ \bibnamefont
  {Mantegna}}\ and\ \bibinfo {author} {\bibfnamefont {H.~E.}\ \bibnamefont
  {Stanley}},\ }\href@noop {} {\emph {\bibinfo {title} {{An Introduction to
  Econophysics: Correlations and Complexity in Finance}}}}\ (\bibinfo
  {publisher} {Cambridge University Press, New York},\ \bibinfo {year}
  {2000})\BibitemShut {NoStop}%
\bibitem [{Note7()}]{Note7}%
  \BibitemOpen
  \bibinfo {note} {\protect \url
  {https://www.investing.com/indices/wig-historical-data}, accessed on November
  10, 2019}\BibitemShut {NoStop}%
\bibitem [{\citenamefont {{Urry}}\ and\ \citenamefont
  {{Padovani}}(1995)}]{Urry95}%
  \BibitemOpen
  \bibfield  {author} {\bibinfo {author} {\bibfnamefont {C.~M.}\ \bibnamefont
  {{Urry}}}\ and\ \bibinfo {author} {\bibfnamefont {P.}~\bibnamefont
  {{Padovani}}},\ }\href {\doibase 10.1086/133630} {\bibfield  {journal}
  {\bibinfo  {journal} {Publications of the Astronomical Society of the
  Pacific}\ }\textbf {\bibinfo {volume} {107}},\ \bibinfo {pages} {803}
  (\bibinfo {year} {1995})}\BibitemShut {NoStop}%
\bibitem [{\citenamefont {{{\.Z}ywucka}}\ \emph {et~al.}(2019)\citenamefont
  {{{\.Z}ywucka}}, \citenamefont {{Tarnopolski}}, \citenamefont
  {{B{\"o}ttcher}}, \citenamefont {{Stawarz}},\ and\ \citenamefont
  {{Marchenko}}}]{zywucka19}%
  \BibitemOpen
  \bibfield  {author} {\bibinfo {author} {\bibfnamefont {N.}~\bibnamefont
  {{{\.Z}ywucka}}}, \bibinfo {author} {\bibfnamefont {M.}~\bibnamefont
  {{Tarnopolski}}}, \bibinfo {author} {\bibfnamefont {M.}~\bibnamefont
  {{B{\"o}ttcher}}}, \bibinfo {author} {\bibfnamefont {{\L}.}~\bibnamefont
  {{Stawarz}}}, \ and\ \bibinfo {author} {\bibfnamefont {V.}~\bibnamefont
  {{Marchenko}}},\ }\href@noop {} {\bibfield  {journal} {\bibinfo  {journal}
  {arXiv e-prints}\ ,\ \bibinfo {eid} {arXiv:1912.03530}} (\bibinfo {year}
  {2019})},\ \Eprint {http://arxiv.org/abs/1912.03530} {arXiv:1912.03530
  [astro-ph.GA]} \BibitemShut {NoStop}%
\bibitem [{\citenamefont {{Zhang}}\ \emph {et~al.}(1993)\citenamefont
  {{Zhang}}, \citenamefont {{Mouritsen}},\ and\ \citenamefont
  {{Zuckermann}}}]{zhang93}%
  \BibitemOpen
  \bibfield  {author} {\bibinfo {author} {\bibfnamefont {Z.}~\bibnamefont
  {{Zhang}}}, \bibinfo {author} {\bibfnamefont {O.~G.}\ \bibnamefont
  {{Mouritsen}}}, \ and\ \bibinfo {author} {\bibfnamefont {M.~J.}\ \bibnamefont
  {{Zuckermann}}},\ }\href {\doibase 10.1103/PhysRevE.48.R2327} {\bibfield
  {journal} {\bibinfo  {journal} {Physical Review E}\ }\textbf {\bibinfo
  {volume} {48}},\ \bibinfo {pages} {R2327} (\bibinfo {year}
  {1993})}\BibitemShut {NoStop}%
\bibitem [{\citenamefont {{Chlouverakis}}\ \emph {et~al.}(2008)\citenamefont
  {{Chlouverakis}}, \citenamefont {{Argyris}}, \citenamefont {{Bogris}},\ and\
  \citenamefont {{Syvridis}}}]{chlouverakis08}%
  \BibitemOpen
  \bibfield  {author} {\bibinfo {author} {\bibfnamefont {K.~E.}\ \bibnamefont
  {{Chlouverakis}}}, \bibinfo {author} {\bibfnamefont {A.}~\bibnamefont
  {{Argyris}}}, \bibinfo {author} {\bibfnamefont {A.}~\bibnamefont {{Bogris}}},
  \ and\ \bibinfo {author} {\bibfnamefont {D.}~\bibnamefont {{Syvridis}}},\
  }\href {\doibase 10.1103/PhysRevE.78.066215} {\bibfield  {journal} {\bibinfo
  {journal} {Physical Review E}\ }\textbf {\bibinfo {volume} {78}},\ \bibinfo
  {eid} {066215} (\bibinfo {year} {2008})}\BibitemShut {NoStop}%
\bibitem [{\citenamefont {{Anzolin}}\ \emph {et~al.}(2010)\citenamefont
  {{Anzolin}}, \citenamefont {{Tamburini}}, \citenamefont {{de Martino}},\ and\
  \citenamefont {{Bianchini}}}]{anzolin10}%
  \BibitemOpen
  \bibfield  {author} {\bibinfo {author} {\bibfnamefont {G.}~\bibnamefont
  {{Anzolin}}}, \bibinfo {author} {\bibfnamefont {F.}~\bibnamefont
  {{Tamburini}}}, \bibinfo {author} {\bibfnamefont {D.}~\bibnamefont {{de
  Martino}}}, \ and\ \bibinfo {author} {\bibfnamefont {A.}~\bibnamefont
  {{Bianchini}}},\ }\href {\doibase 10.1051/0004-6361/201014297} {\bibfield
  {journal} {\bibinfo  {journal} {Astronomy and Astrophysics}\ }\textbf
  {\bibinfo {volume} {519}},\ \bibinfo {eid} {A69} (\bibinfo {year} {2010})},\
  \Eprint {http://arxiv.org/abs/1005.5319} {arXiv:1005.5319 [astro-ph.SR]}
  \BibitemShut {NoStop}%
\bibitem [{\citenamefont {{Kijima}}\ \emph {et~al.}(2014)\citenamefont
  {{Kijima}}, \citenamefont {{Yokoyama}}, \citenamefont {{Shima}},\ and\
  \citenamefont {{Yamamoto}}}]{kijima14}%
  \BibitemOpen
  \bibfield  {author} {\bibinfo {author} {\bibfnamefont {A.}~\bibnamefont
  {{Kijima}}}, \bibinfo {author} {\bibfnamefont {K.}~\bibnamefont
  {{Yokoyama}}}, \bibinfo {author} {\bibfnamefont {H.}~\bibnamefont {{Shima}}},
  \ and\ \bibinfo {author} {\bibfnamefont {Y.}~\bibnamefont {{Yamamoto}}},\
  }\href {\doibase 10.1140/epjb/e2014-40987-5} {\bibfield  {journal} {\bibinfo
  {journal} {European Physical Journal B}\ }\textbf {\bibinfo {volume} {87}},\
  \bibinfo {eid} {41} (\bibinfo {year} {2014})},\ \Eprint
  {http://arxiv.org/abs/1402.0912} {arXiv:1402.0912 [physics.pop-ph]}
  \BibitemShut {NoStop}%
\bibitem [{\citenamefont {{Makarava}}\ \emph {et~al.}(2014)\citenamefont
  {{Makarava}}, \citenamefont {{Menz}}, \citenamefont {{Theves}}, \citenamefont
  {{Huisinga}}, \citenamefont {{Beta}},\ and\ \citenamefont
  {{Holschneider}}}]{makarava14}%
  \BibitemOpen
  \bibfield  {author} {\bibinfo {author} {\bibfnamefont {N.}~\bibnamefont
  {{Makarava}}}, \bibinfo {author} {\bibfnamefont {S.}~\bibnamefont {{Menz}}},
  \bibinfo {author} {\bibfnamefont {M.}~\bibnamefont {{Theves}}}, \bibinfo
  {author} {\bibfnamefont {W.}~\bibnamefont {{Huisinga}}}, \bibinfo {author}
  {\bibfnamefont {C.}~\bibnamefont {{Beta}}}, \ and\ \bibinfo {author}
  {\bibfnamefont {M.}~\bibnamefont {{Holschneider}}},\ }\href {\doibase
  10.1103/PhysRevE.90.042703} {\bibfield  {journal} {\bibinfo  {journal}
  {Physical Review E}\ }\textbf {\bibinfo {volume} {90}},\ \bibinfo {eid}
  {042703} (\bibinfo {year} {2014})}\BibitemShut {NoStop}%
\bibitem [{\citenamefont {{Murgu{\'\i}a}}\ \emph {et~al.}(2015)\citenamefont
  {{Murgu{\'\i}a}}, \citenamefont {{Rosu}}, \citenamefont {{Jimenez}},
  \citenamefont {{Guti{\'e}rrez-Medina}},\ and\ \citenamefont
  {{Garc{\'\i}a-Meza}}}]{marguia15}%
  \BibitemOpen
  \bibfield  {author} {\bibinfo {author} {\bibfnamefont {J.~S.}\ \bibnamefont
  {{Murgu{\'\i}a}}}, \bibinfo {author} {\bibfnamefont {H.~C.}\ \bibnamefont
  {{Rosu}}}, \bibinfo {author} {\bibfnamefont {A.}~\bibnamefont {{Jimenez}}},
  \bibinfo {author} {\bibfnamefont {B.}~\bibnamefont {{Guti{\'e}rrez-Medina}}},
  \ and\ \bibinfo {author} {\bibfnamefont {J.~V.}\ \bibnamefont
  {{Garc{\'\i}a-Meza}}},\ }\href {\doibase 10.1016/j.physa.2014.09.046}
  {\bibfield  {journal} {\bibinfo  {journal} {Physica A: Statistical Mechanics
  and its Applications}\ }\textbf {\bibinfo {volume} {417}},\ \bibinfo {pages}
  {176} (\bibinfo {year} {2015})},\ \Eprint {http://arxiv.org/abs/1410.3135}
  {arXiv:1410.3135 [q-bio.QM]} \BibitemShut {NoStop}%
\bibitem [{\citenamefont {{Carbone}}\ \emph {et~al.}(2018)\citenamefont
  {{Carbone}}, \citenamefont {{Sorriso-Valvo}}, \citenamefont {{Alberti}},
  \citenamefont {{Lepreti}}, \citenamefont {{Chen}}, \citenamefont
  {{N{\v{e}}me{\v{c}}ek}},\ and\ \citenamefont
  {{{\v{S}}afr{\'a}nkov{\'a}}}}]{carbone18}%
  \BibitemOpen
  \bibfield  {author} {\bibinfo {author} {\bibfnamefont {F.}~\bibnamefont
  {{Carbone}}}, \bibinfo {author} {\bibfnamefont {L.}~\bibnamefont
  {{Sorriso-Valvo}}}, \bibinfo {author} {\bibfnamefont {T.}~\bibnamefont
  {{Alberti}}}, \bibinfo {author} {\bibfnamefont {F.}~\bibnamefont
  {{Lepreti}}}, \bibinfo {author} {\bibfnamefont {C.~H.~K.}\ \bibnamefont
  {{Chen}}}, \bibinfo {author} {\bibfnamefont {Z.}~\bibnamefont
  {{N{\v{e}}me{\v{c}}ek}}}, \ and\ \bibinfo {author} {\bibfnamefont
  {J.}~\bibnamefont {{{\v{S}}afr{\'a}nkov{\'a}}}},\ }\href {\doibase
  10.3847/1538-4357/aabcc2} {\bibfield  {journal} {\bibinfo  {journal}
  {Astrophysical Journal}\ }\textbf {\bibinfo {volume} {859}},\ \bibinfo {eid}
  {27} (\bibinfo {year} {2018})},\ \Eprint {http://arxiv.org/abs/1804.02169}
  {arXiv:1804.02169 [physics.space-ph]} \BibitemShut {NoStop}%
\bibitem [{\citenamefont {{Witton}}\ \emph {et~al.}(2019)\citenamefont
  {{Witton}}, \citenamefont {{Sergeyev}}, \citenamefont {{Turitsyna}},
  \citenamefont {{Furlong}}, \citenamefont {{Seri}}, \citenamefont
  {{Brookes}},\ and\ \citenamefont {{Turitsyn}}}]{witton19}%
  \BibitemOpen
  \bibfield  {author} {\bibinfo {author} {\bibfnamefont {C.}~\bibnamefont
  {{Witton}}}, \bibinfo {author} {\bibfnamefont {S.~V.}\ \bibnamefont
  {{Sergeyev}}}, \bibinfo {author} {\bibfnamefont {E.~G.}\ \bibnamefont
  {{Turitsyna}}}, \bibinfo {author} {\bibfnamefont {P.~L.}\ \bibnamefont
  {{Furlong}}}, \bibinfo {author} {\bibfnamefont {S.}~\bibnamefont {{Seri}}},
  \bibinfo {author} {\bibfnamefont {M.}~\bibnamefont {{Brookes}}}, \ and\
  \bibinfo {author} {\bibfnamefont {S.~K.}\ \bibnamefont {{Turitsyn}}},\ }\href
  {\doibase 10.1088/1741-2552/ab225e} {\bibfield  {journal} {\bibinfo
  {journal} {Journal of Neural Engineering}\ }\textbf {\bibinfo {volume}
  {16}},\ \bibinfo {eid} {056019} (\bibinfo {year} {2019})}\BibitemShut
  {NoStop}%
\bibitem [{\citenamefont {{Zunino}}\ \emph {et~al.}(2008)\citenamefont
  {{Zunino}}, \citenamefont {{P{\'e}rez}}, \citenamefont {{Mart{\'\i}n}},
  \citenamefont {{Garavaglia}}, \citenamefont {{Plastino}},\ and\ \citenamefont
  {{Rosso}}}]{zunino08}%
  \BibitemOpen
  \bibfield  {author} {\bibinfo {author} {\bibfnamefont {L.}~\bibnamefont
  {{Zunino}}}, \bibinfo {author} {\bibfnamefont {D.~G.}\ \bibnamefont
  {{P{\'e}rez}}}, \bibinfo {author} {\bibfnamefont {M.~T.}\ \bibnamefont
  {{Mart{\'\i}n}}}, \bibinfo {author} {\bibfnamefont {M.}~\bibnamefont
  {{Garavaglia}}}, \bibinfo {author} {\bibfnamefont {A.}~\bibnamefont
  {{Plastino}}}, \ and\ \bibinfo {author} {\bibfnamefont {O.~A.}\ \bibnamefont
  {{Rosso}}},\ }\href {\doibase 10.1016/j.physleta.2008.05.026} {\bibfield
  {journal} {\bibinfo  {journal} {Physics Letters A}\ }\textbf {\bibinfo
  {volume} {372}},\ \bibinfo {pages} {4768} (\bibinfo {year}
  {2008})}\BibitemShut {NoStop}%
\bibitem [{\citenamefont {Sadhu}\ \emph {et~al.}(2018)\citenamefont {Sadhu},
  \citenamefont {Delorme},\ and\ \citenamefont {Wiese}}]{Sadhu18}%
  \BibitemOpen
  \bibfield  {author} {\bibinfo {author} {\bibfnamefont {T.}~\bibnamefont
  {Sadhu}}, \bibinfo {author} {\bibfnamefont {M.}~\bibnamefont {Delorme}}, \
  and\ \bibinfo {author} {\bibfnamefont {K.~J.}\ \bibnamefont {Wiese}},\ }\href
  {\doibase 10.1103/PhysRevLett.120.040603} {\bibfield  {journal} {\bibinfo
  {journal} {Phys. Rev. Lett.}\ }\textbf {\bibinfo {volume} {120}},\ \bibinfo
  {pages} {040603} (\bibinfo {year} {2018})}\BibitemShut {NoStop}%
\bibitem [{\citenamefont {{Shin}}\ \emph {et~al.}(2009)\citenamefont {{Shin}},
  \citenamefont {{Sekora}},\ and\ \citenamefont {{Byun}}}]{Shin09}%
  \BibitemOpen
  \bibfield  {author} {\bibinfo {author} {\bibfnamefont {M.-S.}\ \bibnamefont
  {{Shin}}}, \bibinfo {author} {\bibfnamefont {M.}~\bibnamefont {{Sekora}}}, \
  and\ \bibinfo {author} {\bibfnamefont {Y.-I.}\ \bibnamefont {{Byun}}},\
  }\href {\doibase 10.1111/j.1365-2966.2009.15576.x} {\bibfield  {journal}
  {\bibinfo  {journal} {Monthly Notices of the Royal Astronomical Society}\
  }\textbf {\bibinfo {volume} {400}},\ \bibinfo {pages} {1897} (\bibinfo {year}
  {2009})}\BibitemShut {NoStop}%
\bibitem [{\citenamefont {{Sokolovsky}}\ \emph {et~al.}(2017)\citenamefont
  {{Sokolovsky}}, \citenamefont {{Gavras}}, \citenamefont {{Karampelas}},
  \citenamefont {{Antipin}}, \citenamefont {{Bellas-Velidis}}, \citenamefont
  {{Benni}}, \citenamefont {{Bonanos}}, \citenamefont {{Burdanov}},
  \citenamefont {{Derlopa}}, \citenamefont {{Hatzidimitriou}}, \citenamefont
  {{Khokhryakova}}, \citenamefont {{Kolesnikova}}, \citenamefont {{Korotkiy}},
  \citenamefont {{Lapukhin}}, \citenamefont {{Moretti}}, \citenamefont
  {{Popov}}, \citenamefont {{Pouliasis}}, \citenamefont {{Samus}},
  \citenamefont {{Spetsieri}}, \citenamefont {{Veselkov}}, \citenamefont
  {{Volkov}}, \citenamefont {{Yang}},\ and\ \citenamefont
  {{Zubareva}}}]{sokolovsky16}%
  \BibitemOpen
  \bibfield  {author} {\bibinfo {author} {\bibfnamefont {K.~V.}\ \bibnamefont
  {{Sokolovsky}}}, \bibinfo {author} {\bibfnamefont {P.}~\bibnamefont
  {{Gavras}}}, \bibinfo {author} {\bibfnamefont {A.}~\bibnamefont
  {{Karampelas}}}, \bibinfo {author} {\bibfnamefont {S.~V.}\ \bibnamefont
  {{Antipin}}}, \bibinfo {author} {\bibfnamefont {I.}~\bibnamefont
  {{Bellas-Velidis}}}, \bibinfo {author} {\bibfnamefont {P.}~\bibnamefont
  {{Benni}}}, \bibinfo {author} {\bibfnamefont {A.~Z.}\ \bibnamefont
  {{Bonanos}}}, \bibinfo {author} {\bibfnamefont {A.~Y.}\ \bibnamefont
  {{Burdanov}}}, \bibinfo {author} {\bibfnamefont {S.}~\bibnamefont
  {{Derlopa}}}, \bibinfo {author} {\bibfnamefont {D.}~\bibnamefont
  {{Hatzidimitriou}}}, \bibinfo {author} {\bibfnamefont {A.~D.}\ \bibnamefont
  {{Khokhryakova}}}, \bibinfo {author} {\bibfnamefont {D.~M.}\ \bibnamefont
  {{Kolesnikova}}}, \bibinfo {author} {\bibfnamefont {S.~A.}\ \bibnamefont
  {{Korotkiy}}}, \bibinfo {author} {\bibfnamefont {E.~G.}\ \bibnamefont
  {{Lapukhin}}}, \bibinfo {author} {\bibfnamefont {M.~I.}\ \bibnamefont
  {{Moretti}}}, \bibinfo {author} {\bibfnamefont {A.~A.}\ \bibnamefont
  {{Popov}}}, \bibinfo {author} {\bibfnamefont {E.}~\bibnamefont
  {{Pouliasis}}}, \bibinfo {author} {\bibfnamefont {N.~N.}\ \bibnamefont
  {{Samus}}}, \bibinfo {author} {\bibfnamefont {Z.}~\bibnamefont
  {{Spetsieri}}}, \bibinfo {author} {\bibfnamefont {S.~A.}\ \bibnamefont
  {{Veselkov}}}, \bibinfo {author} {\bibfnamefont {K.~V.}\ \bibnamefont
  {{Volkov}}}, \bibinfo {author} {\bibfnamefont {M.}~\bibnamefont {{Yang}}}, \
  and\ \bibinfo {author} {\bibfnamefont {A.~M.}\ \bibnamefont {{Zubareva}}},\
  }\href {\doibase 10.1093/mnras/stw2262} {\bibfield  {journal} {\bibinfo
  {journal} {Monthly Notices of the Royal Astronomical Society}\ }\textbf
  {\bibinfo {volume} {464}},\ \bibinfo {pages} {274} (\bibinfo {year}
  {2017})}\BibitemShut {NoStop}%
\bibitem [{\citenamefont {{P{\'e}rez-Ortiz}}\ \emph {et~al.}(2017)\citenamefont
  {{P{\'e}rez-Ortiz}}, \citenamefont {{Garc{\'{\i}}a-Varela}}, \citenamefont
  {{Quiroz}}, \citenamefont {{Sabogal}},\ and\ \citenamefont
  {{Hern{\'a}ndez}}}]{perez17}%
  \BibitemOpen
  \bibfield  {author} {\bibinfo {author} {\bibfnamefont {M.~F.}\ \bibnamefont
  {{P{\'e}rez-Ortiz}}}, \bibinfo {author} {\bibfnamefont {A.}~\bibnamefont
  {{Garc{\'{\i}}a-Varela}}}, \bibinfo {author} {\bibfnamefont {A.~J.}\
  \bibnamefont {{Quiroz}}}, \bibinfo {author} {\bibfnamefont {B.~E.}\
  \bibnamefont {{Sabogal}}}, \ and\ \bibinfo {author} {\bibfnamefont
  {J.}~\bibnamefont {{Hern{\'a}ndez}}},\ }\href {\doibase
  10.1051/0004-6361/201628937} {\bibfield  {journal} {\bibinfo  {journal}
  {Astronomy \& Astrophysics}\ }\textbf {\bibinfo {volume} {605}},\ \bibinfo
  {eid} {A123} (\bibinfo {year} {2017})}\BibitemShut {NoStop}%
\bibitem [{\citenamefont {{Lafler}}\ and\ \citenamefont
  {{Kinman}}(1965)}]{lafler65}%
  \BibitemOpen
  \bibfield  {author} {\bibinfo {author} {\bibfnamefont {J.}~\bibnamefont
  {{Lafler}}}\ and\ \bibinfo {author} {\bibfnamefont {T.~D.}\ \bibnamefont
  {{Kinman}}},\ }\href {\doibase 10.1086/190116} {\bibfield  {journal}
  {\bibinfo  {journal} {Astrophysical Journal Supplement}\ }\textbf {\bibinfo
  {volume} {11}},\ \bibinfo {pages} {216} (\bibinfo {year} {1965})}\BibitemShut
  {NoStop}%
\bibitem [{\citenamefont {{Zanin}}\ \emph {et~al.}(2012)\citenamefont
  {{Zanin}}, \citenamefont {{Zunino}}, \citenamefont {{Rosso}},\ and\
  \citenamefont {{Papo}}}]{zanin12}%
  \BibitemOpen
  \bibfield  {author} {\bibinfo {author} {\bibfnamefont {M.}~\bibnamefont
  {{Zanin}}}, \bibinfo {author} {\bibfnamefont {L.}~\bibnamefont {{Zunino}}},
  \bibinfo {author} {\bibfnamefont {O.~A.}\ \bibnamefont {{Rosso}}}, \ and\
  \bibinfo {author} {\bibfnamefont {D.}~\bibnamefont {{Papo}}},\ }\href
  {\doibase 10.3390/e14081553} {\bibfield  {journal} {\bibinfo  {journal}
  {Entropy}\ }\textbf {\bibinfo {volume} {14}},\ \bibinfo {pages} {1553}
  (\bibinfo {year} {2012})}\BibitemShut {NoStop}%
\bibitem [{\citenamefont {Brockwell}(2014)}]{Brockwell2014}%
  \BibitemOpen
  \bibfield  {author} {\bibinfo {author} {\bibfnamefont {P.~J.}\ \bibnamefont
  {Brockwell}},\ }\href {\doibase 10.1007/s10463-014-0468-7} {\bibfield
  {journal} {\bibinfo  {journal} {Annals of the Institute of Statistical
  Mathematics}\ }\textbf {\bibinfo {volume} {66}},\ \bibinfo {pages} {647}
  (\bibinfo {year} {2014})}\BibitemShut {NoStop}%
\bibitem [{\citenamefont {Tsai}(2009)}]{Tsai09}%
  \BibitemOpen
  \bibfield  {author} {\bibinfo {author} {\bibfnamefont {H.}~\bibnamefont
  {Tsai}},\ }\href@noop {} {\bibfield  {journal} {\bibinfo  {journal}
  {Bernoulli}\ }\textbf {\bibinfo {volume} {15}},\ \bibinfo {pages} {178}
  (\bibinfo {year} {2009})}\BibitemShut {NoStop}%
\end{thebibliography}%

\end{document}